\documentclass[twocolumn,superscriptaddress]{aastex7}
\usepackage{graphicx}
\usepackage{subfigure}
\usepackage{subcaption}
\usepackage{bm}
\usepackage{xcolor}
\usepackage{float}
\usepackage{amsmath}
\usepackage{ulem}

\begin{document}

\title{ Results from an Einstein@Home search for continuous gravitational waves from Cassiopeia A and Vela Jr. using LIGO O2 data}

\author{Jorge Morales}
\email{jormoral@iu.edu}
\affiliation{ Center for the Exploration of Energy and Matter and Department of Physics, Indiana University, Bloomington, IN 47405, USA }
\affiliation{ Physics Department, Indiana University, Bloomington, IN, 47405, USA } 
\affiliation{ Max Planck Institute for Gravitational Physics (Albert Einstein Institute), Callinstrasse 38, 30167, Hannover, Germany }

\author{Jing Ming}
\email{jing.ming@aei.mpg.de}
\affiliation{ Max Planck Institute for Gravitational Physics (Albert Einstein Institute), Callinstrasse 38, 30167, Hannover, Germany }
\affiliation{ Leibniz Universit{\"a}t Hannover, D-30167 Hannover, Germany }

\author{Maria Alessandra Papa}
\email{maria.alessandra.papa@aei.mpg.de}
\affiliation{ Max Planck Institute for Gravitational Physics (Albert Einstein Institute), Callinstrasse 38, 30167, Hannover, Germany }
\affiliation{ Leibniz Universit{\"a}t Hannover, D-30167 Hannover, Germany }

\author{Heinz-Bernd Eggenstein}
\email{heinz-bernd.eggenstein@aei.mpg.de}
\affiliation{ Max Planck Institute for Gravitational Physics (Albert Einstein Institute), Callinstrasse 38, 30167, Hannover, Germany }
\affiliation{ Leibniz Universit{\"a}t Hannover, D-30167 Hannover, Germany }

\author{Bernd Machenschalk}
\email{Bernd.Machenschalk@aei.mpg.de}
\affiliation{ Max Planck Institute for Gravitational Physics (Albert Einstein Institute), Callinstrasse 38, 30167, Hannover, Germany }
\affiliation{ Leibniz Universit{\"a}t Hannover, D-30167 Hannover, Germany }

\correspondingauthor{Jing Ming}
\email{jing.ming@aei.mpg.de}


\begin{abstract}
We conduct two searches for continuous, nearly monochromatic gravitational waves originating from the central compact objects in the supernova remnants Cassiopeia A and Vela Jr. using public LIGO data. The search for Cassiopeia A targets signal frequencies between 20 Hz and 400 Hz; the Vela Jr. search between 400 Hz and 1700 Hz, and both investigate the broadest set of waveforms ever considered with highly sensitive deterministic search methods.
Above 1500 Hz the Vela Jr. search is the most sensitive carried out thus far, improving on previous results by over 300\%. Above 976 Hz these results improve on existing ones by 50\%. In all we investigate over $10^{18}$ waveforms, leveraging the computational power donated by thousands of Einstein@Home volunteers. We perform a 4-stage follow-up on more than 6 million waveforms. None of the considered waveforms survives the follow-up scrutiny, indicating no significant detection candidate.  Our null results constrain the maximum amplitude of continuous signals as a function of signal frequency from the targets. The most stringent 90\% confidence upper limit for Cas A is $h_0^{90 \%}\approx 7.3\times10^{-26}$ near 200 Hz, and for Vela Jr. it is $h_0^{90 \%}\approx 8.9\times10^{-26}$ near 400 Hz. Translated into upper limits on the ellipticity and r-mode amplitude, our results probe physically interesting regions: for example the ellipticity of Vela Jr. is constrained to be smaller than $10^{-7}$ across the frequency band, with a tighter constraint of less than $2\times10^{-8}$ at the highest frequencies. 

\end{abstract}


\section{\label{sec:intro} Introduction}

Continuous gravitational waves are weak and persistent 
quasi-monochromatic signals that remain undetected. Continuous waves are believed to be orders of magnitude weaker than transient gravitational waves resulting from mergers of binary systems of black holes and/or neutron stars. This inherent weakness poses a significant challenge in their detection. On the other hand, continuous waves are thought to persist for years so the data can be integrated over many months against the expected signal templates to amplify the signal-to-noise ratio  and enhance our ability to detect the signal.

Broadly speaking, there are three types of continuous wave searches. Targeted 
searches focus on continuous waves from pulsars with known electromagnetic spin frequency and its evolution \citep{Abbott2019,Abbott2019_2,Abbott2022,Abbott2022_2,Nieder:2020yqy,Rajbhandari2021,Clark:2022tjf,Owen:2023maw,Ashok2021,Ashok2024,Mirasola:2024kll,lvk2025a}. These are the most sensitive and least computationally expensive continuous wave searches. In contrast, all-sky searches aim for continuous waves from unknown sources across a significant portion of the sky \citep{Dergachev2020,Dergachev2021,Singh:2022hfd,Covas2022,covas2024,Abbott2022_3,Steltner2023,Dergachev2023}. The broad parameter space that all-sky searches need to cover makes them the least sensitive and most computationally expensive. Somewhere in between, directed searches look for continuous waves from sources at compelling sky locations where no electromagnetic pulsations have been detected \citep{Abbott2019_3,LVC_O2_V,Dergachev2019,Ming2019,Papa_2020midth,Piccinni2020,Millhouse2020,Lindblom2020,Zhang2021,LIGOScientific:2021mwx,Abbott2022_4,KAGRA:2022osp,LIGOScientific:2022enz,Owen2022,Ming2022,PhysRevD.110.042006,ming2024a}. Given the absence of prior knowledge about the frequency and its evolution, directed searches require convolving the data with a vast array of signal waveforms, incurring a high computational cost.

Continuous waves can arise from dynamics deviating from axial symmetry within
the interior of rotating neutron stars (where the axis of symmetry is the rotation axis). These dynamics include sustained deformations in the neutron star crust, informally known as ``mountains" \citep{Ushomirsky2000,Haskell2006,Gittins2021,Gittins2021_2,Hutchins2022,Morales2022,Morales2024}, and periodic non-radial oscillations of dense fluids within neutron stars, referred to as r-modes or Rossby waves \citep{Andersson1999,Brown2000,Haskell2014,Gittins2023}. Additionally, continuous waves may be generated through more exotic mechanisms, like the rapid inspiral of two compact dark-matter objects \citep{Horowitz2019,Horowitz2020,Miller:2024fpo} and the super-radiant emission of axion-like particles around spinning black holes \citep{Arvanitaki2015,Zhu2020,LIGOScientific:2021rnv,Mirasola:2025car}.

In the current study, we perform directed searches on Cassiopeia 
A (G111.7-2.1) and Vela Jr. (G266.2-1.2). These are two galactic compact objects that reside at the center of young and nearby supernova remnants. These central compact objects are presumed rapidly spinning, isolated neutron stars that harbor most of the angular momentum of their respective progenitors. If these central objects are indeed rapidly rotating neutron stars with significant mountains or r-mode oscillations, a fraction of the substantial angular momentum reservoir is being radiated as continuous waves. 

We implement our searches within the Einstein@Home volunteer distributed computing project \citep{E@H}, with a computing budget spanning several months. The selection of our two targets and their search setups results from an optimization scheme that identifies the sources and search setups that maximize the detection probability within a specified computational budget \citep{Ming2016,Ming2018}. This scheme incorporates factors such as the sensitivity and computing cost of the search pipeline, the age and distance of a source, and priors on the source's frequency, frequency derivatives, and ellipticity. The search sensitivity depends on the search setup and on the noise level in detectors' data. We conduct the post-processing of the results of the Einstein@Home searches using the in-house Atlas super-computing cluster \citep{Atlas} to further enhance sensitivity. 

\section{\label{sec:targets} Targets}
\subsection{\label{subsec:cas_a} Cassiopeia A (G111.7-2.1)}

Cassiopeia A (Cas A) is the central compact object resulting from 
one of the most recent galactic core-collapse supernovae. It is at the center of the supernova remnant CXOU J232327.9+584842, and its position was measured using Chandra X-ray satellite data \citep{Tanabaum1999}. The prevailing consensus is that this explosive event occurred 310-350 years ago \citep{Fesen2006} at a distance of 3.3-3.7 kpc from Earth \citep{Reed1995}. The age and distance are calculated using the observed expansion rate of the outer ejecta of the remnant and the radial velocity of its central compact object. Based on its  X-ray spectrum, \cite{Ho2009} suggest that the central object is neutron star with a carbon atmosphere and a small magnetic field. The intricate and asymmetric nebula surrounding Cas A is a potential fingerprint of an asymmetric neutron star interior. 
Additionally, if the newly born neutron star was fast-spinning and with a relatively weak internal magnetic field as \cite{Ho2009} suggest, it might well be subject to rotational instabilities like r-modes \citep{Owen:1998xg}.

\subsection{\label{subsec:vela_jr} Vela Jr. (G266.2-1.2)}

The X-ray-to-optical-flux ratio reveals that Vela Jr. is the central compact object of the supernova remnant CXOU J085201.4- 
461753. 
Its position was originally pin-pointed by the Chandra X-ray satellite \citep{Pavlov2001}, with subsequent infra-red observations confirming the position and suggesting that the central compact object is a neutron star \citep{Mignani2007}. 

Based on Ti$^{44}$ line emission from the supernova remnant the age of the remnant is estimated to be 700 yrs and its distance to be 200 pc \citep{Iyudin1998}. Conversely Chandra X-ray data and hydrodynamic analyses of the expansion of the remnant indicate an older and more distant object, with an age of 4300 yrs and a distance of 750 pc \citep{Allen2015}. 

\section{\label{sec:signal_model} The Signal Model}

For any plane gravitational wave, there exists a frame that satisfies three key conditions: (1) it is stationary relative to its source, (2) it is perpendicular to the direction of propagation, and (3) it lies parallel to the polarization plane of the wave. In this frame, the gravitational-wave strain can be expressed as a linear combination of two polarizations, denoted as $h_{+}$ and $h_{\times}$:
\begin{equation}
h_+ (t) = A_+ \cos \Phi (t),
\end{equation}
\begin{equation}
h_{\times} (t) = A_{\times} \sin \Phi (t).
\end{equation}
$\Phi (t)$ is the gravitational-wave phase and $A_{+,\times}$ denote the gravitational-wave polarization amplitudes:
\begin{equation}
A_{+} = \frac{1}{2} h_0 (1 + \cos^2 \iota),
\end{equation}
\begin{equation}
A_{\times} = h_0 \cos \iota, 
\end{equation}
where $\iota$ represents the angle between the angular momentum of the neutron star and the line of sight from Earth, while $h_0$ is the intrinsic continuous wave amplitude. 

The frequency of a continuous wave emitted by an isolated rapidly rotating neutron star evolves gradually over time, and its temporal behaviour can be described using a Taylor expansion around a reference time in the Solar System Barycenter (SSB) $\tau_0$:
\begin{equation}
f(\tau) = f_0 + \dot{f}_0 (\tau - \tau_0) + \frac{1}{2} \ddot{f}_0 (\tau - \tau_0)^2.
\end{equation}

The continuous wave strain at the detector takes the form 
\begin{equation}
h(t) = F_+(\alpha,\delta,\psi;t) h_+ (t) + F_{\times}(\alpha,\beta,\psi;t) h_{\times} (t),
\end{equation}
where $F_{+,\times}(\alpha,\delta,\psi;t)$ are the beam-pattern functions, $\alpha$ and $\delta$ represent the right ascension and declination of the source, $\psi$ is for  the orientation angle of the wave-frame with respect to the detector frame, and $t$ denotes the time in the detector frame. The conversion from the detector frame time $t$ to the SSB frame time $\tau$ is determined by
\begin{equation}
\tau (t) = t + \frac{\Vec{r} (t) \cdot \hat{n}}{c} + \Delta_{E \odot} - \Delta_{S \odot},
\end{equation}
where $\Vec{r}$ is the detector position vector, $\hat{n}$ is the unit vector pointing towards the source from the SSB, and $\Delta_{E \odot}$ and $\Delta_{S \odot}$ are the Einstein and Shapiro time delays, respectively.

\section{\label{sec:data} The Data}

In the Einstein@Home searches, we analyze public data obtained during the second observing run (O2) of the two LIGO detectors, taken between GPS times 1167993370 (January 9, 2017) and 1187731774 (August 25, 2017) \citep{Abbott:2021boh}. During this period, the Hanford, Washington detector exhibited a duty cycle of 65 \%, while the Livingston, Louisiana detector had a duty cycle of 62\%. In the post-processing of the Einstein@Home search results, we use public data obtained during the first half of the third observing run (O3a) of the two LIGO detectors, taken between GPS times 1238421231 (April 04 2019) and 1253973231 (October 01 2019) \citep{Abbott_2023}. Duty factors of detectors of Hanford and Livingston during this period are 71\% for  and 76\% respectively.  
The O2 and O3a search reference times we adopt are $\tau_{\textrm{SSB}}^{\textrm{O2}}=1177858472.0$ and   $\tau_{\textrm{SSB}}^{\textrm{O3a}}=1246197626.5$.
We generate Short Fourier Transforms of data segments, each spanning 1800 seconds. We eliminate data affected by artifacts such as calibration lines and main power lines, as well as loud transient glitches \citep{Steltner2022}.

\section{\label{sec:search} The Einstein@Home Search}

We employ a ``stack-slide" approach for semi-coherent searches \citep{Brady:1997ji,2000PhRvD..61h2001B},leveraging the Global Correlation Transform (GCT) method \citep{Pletsch2008,Pletsch2009,Pletsch2010}. In this type of search, the dataset, which spans a total observing time $T_{\text{obs}}$, is partitioned into shorter $N_{\text{seg}}$ segments of equal span $T_{\text{coh}}$. The data in each segment $i$ is analyzed coherently. Afterwards, the coherent detection statistics ${\mathcal{F}}_i$ from all the segments are summed. The coherent detection statistic from each segment is a matched filter between the data and a waveform model with parameters \{$f$, $\dot{f}$, $\ddot{f}$, $\alpha$, $\delta$\}. 
The final detection statistic is
\begin{equation}
   \Bar{\mathcal{F}} = \frac{1}{N_\mathrm{seg}} \sum_{i=1}^{N_\mathrm{seg}} \mathcal{F}_i\, .
   \label{eqn:mean_2F}
\end{equation}

In the ideal scenario of Gaussian noise, $ N_\mathrm{seg}\Bar{\mathcal{F}}$ follows a chi-squared distribution with $4 N_{\text{seg}}$ degrees of freedom and a non-centrality parameter $\rho^2=\sum_i\rho_i^2$, the sum of the squared signal-to-noise ratios from the individual segments. In each segment $\rho_i^2\propto \frac{h_0^2 T_{\text{coh}}}{S_h}$, where $S_h$ is the strain power spectral density of the noise at the signal's frequency \citep{JKS1998}. 

While many disturbances are mitigated with the data preparation techniques  \citep{Steltner2022}, some spectral lines persist. These residual lines can elevate the values of $\Bar{\mathcal{F}}$ for search templates that use data affected by those disturbances. For this reason, a line robust detection statistic $\hat{\beta}_{\text{S/GLtL}}$ is used to rank the Einstein@Home search results. This detection statistic is the log of a Bayesian odds ratio that tests the signal (S) hypothesis versus an extended noise hypothesis. The extended noise hypothesis considers Gaussian (G) noise or line-noise (L) or  transient-line noise (tL)
\citep{Keitel:2013wga, Keitel:2015ova}. The usage of $\hat{\beta}_{\text{S/GLtL}}$ as a ranking detection statistic reduces the incidences of false alarms caused by noise that resembles a continuous wave more than it resembles Gaussian noise.

The search setup consists of the coherent baseline time $T_\mathrm{coh}$, the signal-parameter ranges and the grid-spacings of the templates $\delta f$, $\delta \dot{f}$ and $\delta \ddot{f}$. 
For each of the searches, the grid-spacings in frequency and spindown parameters over the parameter space are constant. Every search setup has an associated average mismatch $\Bar{m}$, which measures the average loss in signal-to-noise ratio due to the mismatch between the parameters of a signal and the closest ($f$, $\dot{f}$ and $\ddot{f}$) grid point. 
The search setups are given at the top of Table \ref{tab:search_params_FU}. 

The search ranges of the spin-down parameters are frequency-dependent, as first proposed by \cite{wette2008}.
We use different search ranges for Cas A and Vela Jr.:
\begin{equation}
    \begin{cases}
         20 \ \text{Hz} & \leq  f\leq 400 \ \text{Hz} ~~~~~~{\textrm{for Cas A}}\\
        400 \ \text{Hz} & \leq  f \leq 1700 \ \text{Hz}~~~~{\textrm{for Vela Jr.}} \\
         -f/ \tau & \leq  \dot{f} \leq 0 \ \text{Hz/s} \\
         0 \ \text{Hz/s$^2$} & \leq \ddot{f}\leq 7 \ f/ {\tau}^2 \\
    \end{cases}
    \label{eqn:param_space}
\end{equation}
where $\tau=330$ years and $\tau=700$ years are the ages of Cas A and Vela Jr., respectively. The choice of the 700 years age for Vela Jr. produces the 
broadest $\dot{f}$ and $\ddot{f}$ search ranges, and it is hence the safest choice. The ranges for $\dot{f}$ and $\ddot{f}$ defined above encompass waveforms from all emissions mechanisms with braking index $n\leq 7$, and more. In fact Eq.s~\ref{eqn:param_space} further expand the template bank in two ways: 1) extending the ranges of $\dot{f}$ and $\ddot{f}$ to 0,  2) not requiring consistency between the underlying braking index of the $\dot{f}$ and $\ddot{f}$ values. We make this choice because it is easy to implement within the search pipeline and because indications exist that a strict power law frequency evolution, especially in young pulsars, may not be completely adequate to describe the observations -- see for instance \cite{Vargas:2024itq} and references therein.

\addtolength{\tabcolsep}{1pt} 
\begin{table*}[t]
    \begin{tabular}{|l c c c c c c|}
        \hline\hline
        Stage & $T_{\text{coh}}$ & $N_{\text{seg}}$ & $\delta f$ & $\delta \dot{f}$  & $\delta \ddot{f}$ & $\Bar{m}$ \\ 
        \hline\hline
         
        E@H (Cas A) & $240$ hr & $17$ & $4.9 \times 10^{-7}$ \ Hz & $8.3 \times 10^{-14}$ Hz/s & $3.4 \times 10^{-20}$ Hz/s$^2$ & 14.3\% \\
        E@H (Vela Jr.) & $240$ hr & $17$ & $4.0 \times 10^{-7}$ Hz & $8.3 \times 10^{-14}$ Hz/s  & $2.7 \times 10^{-20}$ Hz/s$^2$ & 10.5\%\\ 
        \hline
        \hline
        1 & $1080$ hr & $5$ & $1.3 \times 10^{-7}$ \ Hz & $1.5 \times 10^{-14}$ Hz/s  & $1.2 \times 10^{-20}$ Hz/s$^2$ &  $13.3 \ \%$ \\
        2 & $2760$ hr & $2$ & $3.5 \times 10^{-8}$ \ Hz & $2.2 \times 10^{-15}$ Hz/s & $1.2\times 10^{-21}$ Hz/s$^2$ &   $7.6 \ \%$ \\
        3 & $5496$ hr & $1$ & $9.6 \times 10^{-9}$ \ Hz & $2.9 \times 10^{-15}$ Hz/s  & $1.5 \times 10^{-21}$ Hz/s$^2$  &  $2.0 \ \%$ \\ 
        4 (new data)  & $1440$ hr & $3$ & $6.7 \times 10^{-8}$ Hz & $1.2 \times 10^{-14}$ Hz/s  & $1.6 \times 10^{-21}$ Hz/s$^2$ & $5.4 \ \%$ \\
        \hline
        \hline
    \end{tabular}
    \caption{\label{tab:search_params_FU} Search setups used. The Einstein@Home searches ( ``E@H" ) are different for the two targets whereas the follow-up Stages 1-4 use the same setup for both targets. $T_{\text{coh}}$ is the coherent baseline time, $N_{\text{seg}}$ is the number of coherent segments, $\delta f$, $\delta \dot{f}$, and $\delta \ddot{f}$ are the grid-spacings of the templates, and $\Bar{m}$ is the average mismatch. }
\end{table*}
\addtolength{\tabcolsep}{1.5pt} 

\addtolength{\tabcolsep}{1pt} 
\begin{table*}[t]
    \begin{tabular}{|| l  c c c c c c  ||}
        \hline
        \hline
        \multicolumn{7}{||c||}{\bf{Cas A}} \\
         \hline\hline
        Stage & $\Delta f$ [Hz] & $\Delta \dot{f}$ [Hz/s] & $\Delta \ddot{f}$ [Hz/s$^2$] &  $R^a_{\text{thr}}$ & $N_{\text{in}}$ & $N_{\text{out}}$ \\ 
        \hline\hline
        E@H &  full range  & full range & full range  & - & $1.0\times10^{17}$  & $1,516,000$ \\
        1 & $1.2 \times 10^{-6}$  & $1.4 \times 10^{-13}$  & $4.2 \times 10^{-20}$ & 1.8 & $1,516,000$   & $801,589$  \\
        2 &  $3.0 \times 10^{-7}$  & $3.3 \times 10^{-14}$ & $1.1 \times 10^{-20}$  & $4.9$ & $801,589$  & $102,661$  \\
        3 & $6.5 \times 10^{-8}$ & $4.4 \times 10^{-15}$ & $3.3 \times 10^{-21}$  & $9.0$ & $102,661$ & $45,483$  \\
        4 (new data) &  $> 3.2 \times 10^{-8}$  & $> 4.8 \times 10^{-15}$ & $3.1 \times 10^{-21}$  & 3.5 & $45,483$  & $0$ \\
        \hline
\end{tabular}
    
        \begin{tabular}{|| l c c c c c c ||}
        \hline
        \multicolumn{7}{||c||}{\bf{Vela Jr. }} \\
         \hline\hline
        Stage & $\Delta f$ [Hz] & $\Delta \dot{f}$ [Hz/s] & $\Delta \ddot{f}$ [Hz/s$^2$] &  $R^a_{\text{thr}}$ & $N_{\text{in}}$ & $N_{\text{out}}$ \\ 
                E@H &  full range  & full range & full range  & - & $1.2\times10^{18}$  & $5,199,849$\\
        1 & $6.3 \times 10^{-6}$  & $6.1 \times 10^{-14}$  & $2.3 \times 10^{-20}$ & 1.9 & $5,199,849$   & $1,822,369$  \\
        2 &  $2.9 \times 10^{-7}$ & $9.3 \times 10^{-14}$  & $1.3 \times 10^{-20}$  & $5.0$ & $1,822,369$   & $202,655$   \\
        3 & $5.3 \times 10^{-8}$ & $5.9 \times 10^{-14}$ & $3.4 \times 10^{-21}$  & $9.4$ & $202,655$ & $70,259$ \\
        4 (new data) &  $> 3.0 \times 10^{-8}$  & $> 4.7 \times 10^{-15}$ & $2.9 \times 10^{-21}$  & 3.1 & $70,259$ & $0$ \\
        \hline
       \hline
    \end{tabular}
    \caption{\label{tab:results_FU} $\Delta f$, $\Delta \dot{f}$, and $\Delta \ddot{f}$ indicate the extent of the search regions per candidate for Stages 1-4 and for the original Einstein@Home (E@H) search.  For Stage 4 the extents represent the non-evolved uncertainties, since the evolved ones vary depending on the parameters of the specific candidate. $N_{\text{in}}$ represents the number of candidates searched for Stages 1-4 and the total number of templates for the E@H search. $N_{\text{out}}$ is the number of surviving candidates from each stage. $R^a_\text{thr}$ is the threshold of Eq.~\ref{eqn:R_athr} for Stages a=1-4. The results from the E@H search are selected based on the pixeling procedure described in Section \ref{subsec:FU_stage_0} and hence the $R_\textrm{thr}$ value is irrelevant.}
\end{table*}
\addtolength{\tabcolsep}{1.5pt}

The two searches are deployed on the Einstein@Home volunteer computing project, which is built on the BOINC (Berkeley Open Infrastructure for Network Computing) architecture \citep{Anderson2004,Anderson2006}. Einstein@Home uses the idle time of volunteer computer to search for weak astrophysical signals from neutron stars, including continuous waves. 
In total among the two targets, $\approx 1.3\times10^{18}$ waveform templates are searched. 92\% of the templates are used in the search for Vela Jr., because of the broad frequency range. 
The total workload is divided into work-units whose size is chosen to keep the volunteer computer busy for $\approx 8$ CPU hours, resulting in a grand total of 5.7 million work units. 

\begin{figure}[h!tbp]
  \includegraphics[width=\columnwidth]{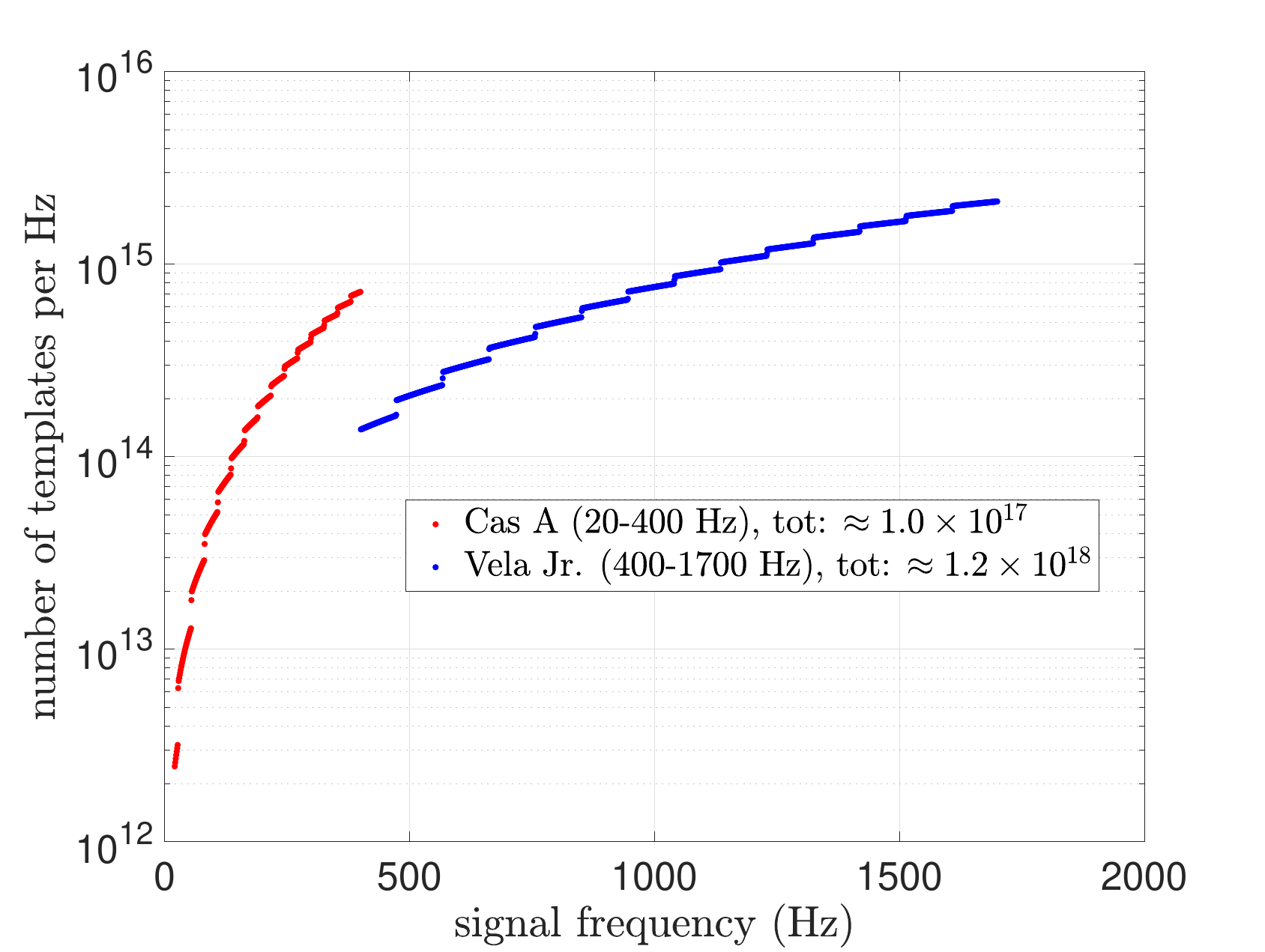}
\caption{Number of templates searched in 1-Hz bands as a function of signal frequency. In the legend we also show the total number of templates searched for each search.}
\label{fig:HowManyTemplates}
\end{figure}

The number of templates searched at a given frequency increases with frequency due to the frequency-dependent search ranges of $\dot{f}$ and $\ddot{f}$  (Eq.s~\ref{eqn:param_space}). Figure \ref{fig:HowManyTemplates} shows the number of templates searched in 1-Hz bands as a function of frequency for the two targets.

\section{Post-processing of Einstein@Home Results}

\subsection{Overview}
\label{sec:overview}

Each Einstein@Home work-unit returns the 10\,000 most significant results. The 5.7 million work units hence return in all $57$ billion candidates.
 
The significance of results calculated on the volunteer computers is expressed by the value of the $\hat{\beta}_{\text{S/GLtL}}$ detection statistic. The larger this value is, the most likely it is that the candidate is associated with an astrophysical signal. The values of $\hat{\beta}_{\text{S/GLtL}}$ at a certain parameter grid point are actually approximations. 
This approximation reduces the computational cost of Einstein@Home searches significantly, but incurs a loss. For the most significant work-unit results the non-approximated value is computed. 
 We indicate the re-computed detection statistics with a  subscript ``r'' as in ``re-computed'': $\hat{\beta}_{\text{S/GLtLr}}$ and $\Bar{\mathcal{F}}_r$. The value of the recomputed statistics is expected to increase in comparison to the non-recomputed value if the waveform template of a candidate is associated to a continuous-wave signal within the data.

Once the detection statistics are re-computed for the $5.7 \times 10^{10}$ most significant Einstein@Home results, we proceed to select the ones that we will follow-up using the \textit{pixeling procedure} described in section \ref{subsec:FU_stage_0}. These constitute the ``Stage-0 candidates".

 The central idea of the follow-up searches is to increase the length of the coherent segments to accumulate more signal-to-noise ratio for candidates associated with astrophysical signals. We define 
\begin{equation}
     R^a \equiv \frac{2 \Bar{\mathcal{F}}_r^{\text{ Stage-a}} - 4}{2 \Bar{\mathcal{F}}_r^{\text{ Stage-0}} - 4} ~~~{\textrm{for a }}=1\cdots 4
     \label{eqn:R_a}
 \end{equation}
which quantifies how much the excess in squared signal-to-noise ratio estimated from the Stage-a candidate ($2 \Bar{\mathcal{F}}_r^{\text{ Stage-a}} - 4$), compares to the excess estimated for the same candidate at Stage-0. Since the square of  signal-to-noise ratio of a signal is roughly proportional to the coherent baseline time $T_{\text{coh}}$ of the data segments, if $T_{\text{coh}}$ increases, it is expected that $R^a$ will also increase. We set a threshold value  $R^a_\mathrm{thr}$  and  
\begin{equation}
     {\textrm{if }} R^a < R^a_\mathrm{thr}  \rightarrow {\textrm{candidate discarded}} .
     \label{eqn:R_athr}
 \end{equation}
Similar to \cite{ming2024a}, $R^a_\mathrm{thr}$  is determined by requiring that it is safe against discarding signals from our target population (see next Section). So thousands of test signals from the target population are added to the data, each is searched for, exactly as done for the real search, and the corresponding $R^a$ is measured. The threshold dismisses at each stage less than $0.01 \%$ of the test signals, resulting in an overall false dismissal of less than $0.04 \%$.

The candidates from this recovery study of signals from the target population also serve a crucial role in determining the region around candidates that must be searched at each stage. 
This region has been known in the literature as ``containment region''. The containment region is defined by a triplet of values at each stage, $\Delta f^a$, $\Delta \dot{f}^a$, and $\Delta \ddot{f}^a$, such that candidates from the target population at that stage are recovered within this distance from the real parameter values. The containment region around the candidates' parameters at Stage-a is the search region at Stage-a+1. 
Tables \ref{tab:results_FU} list the search regions at each stage.

\subsection{Target population}
\label{sec:TargetPop}

The post-processing procedures that follow the Einstein@Home search have been designed to have a very low false dismissal for signals of the target population. The target population signals for a given supernova remnant obviously all come from the position of that source and have frequency and frequency derivative values in the target range, randomly displaced from the search grid points. The $\cos\iota$ and $\psi$ parameters are uniformly distributed $-1\leq \cos\iota \leq 1$, $-\pi/4 \leq\psi \leq \pi/4$. The signal strength is set to the target strength with respect to the amplitude spectral density: the target population signals have amplitudes corresponding to $\mathcal{D}=90 ~[1/\sqrt{\textrm{Hz}}]$ for both Vela Jr. and Cas A (see Eq.~\ref{eqn:sensitivity_depth} for the relation between the amplitude $h_0$ and $\mathcal{D}$). The target depth value is the best (i.e. highest sensitivity) that we could achieve with the available computational power, and in absolute terms it is the highest achieved on such a large parameter space.

\subsection{\label{subsec:FU_stage_0} Stage-0}

The goal at this stage is to efficiently select the most compelling candidates from the $5.7 \times 10^{10}$ results returned to the Einstein@Home server. Historically, there have been three main ways to achieve this.

The first method involves setting a desired threshold on the detection statistic and selecting the search candidates whose detection statistic lies above that threshold. This approach is straightforward and time-saving. However the number of selected candidates from disturbed frequency bands may greatly exceed those in undisturbed bands leading to a computationally wasteful and less effective follow-up process.

The second method is to select a fixed number of candidates per frequency band. The fixed number of selected candidates per band is determined by the available computational budget. We call this method to select follow-up candidates method ``simple thresholding''. In each band, this method selects a fixed amount of candidates, with the quietest candidate setting the detection statistic threshold. Unfortunately, in disturbed bands, this procedure might only select candidates associated with disturbances and  discard interesting candidates not linked to the disturbance.

The third method is called \textit{clustering}. In clustering, candidates associated with the same astrophysical or non-astrophysical causes are identified, and one of them is chosen \citep{Singh:2017kss,den_cluster}. Clustering is generally efficient because it prevents the waste of resources on following up candidates originatign from the same root cause. While clustering has proven to be very effective in all-sky searches, in directed searches, the sensitivity improvement compared with simple thresholding is very modest -- less than $1$ \% \citep{ming2024a}. Since determining a good operating point for the clustering procedure requires time and extensive Monte Carlos, we do not pursue it here. Instead, we propose an alternative scheme to select the Stage-0 results to follow-up: the pixeling procedure. This alternative is straightforward and efficient and does not loose detection efficiency in disturbed frequency bands.


\subsubsection{\label{subsubsec:pixeling_procedure} The Pixeling Procedure}

We divide the searched frequency range in $N_{\textrm{bands}}$ bands of equal size.  For each frequency band, we consider $N_p$ frequency-spindown ``pixels'', with each pixel being a unique small region of the searched parameter space. The frequency range covered by each pixel is 5 mHz. The $\dot{f}$ range covered by a pixel varies with frequency and with the age $\tau$ of the source, and  it is equal to $f/10\tau$.  So we consider 0.5 Hz wide frequency bands, and $N_p=100\times 10$ pixels to cover the $f\times \dot{f}$ parameter space. 

\begin{figure}[tb]
    \includegraphics[width=0.90\columnwidth] {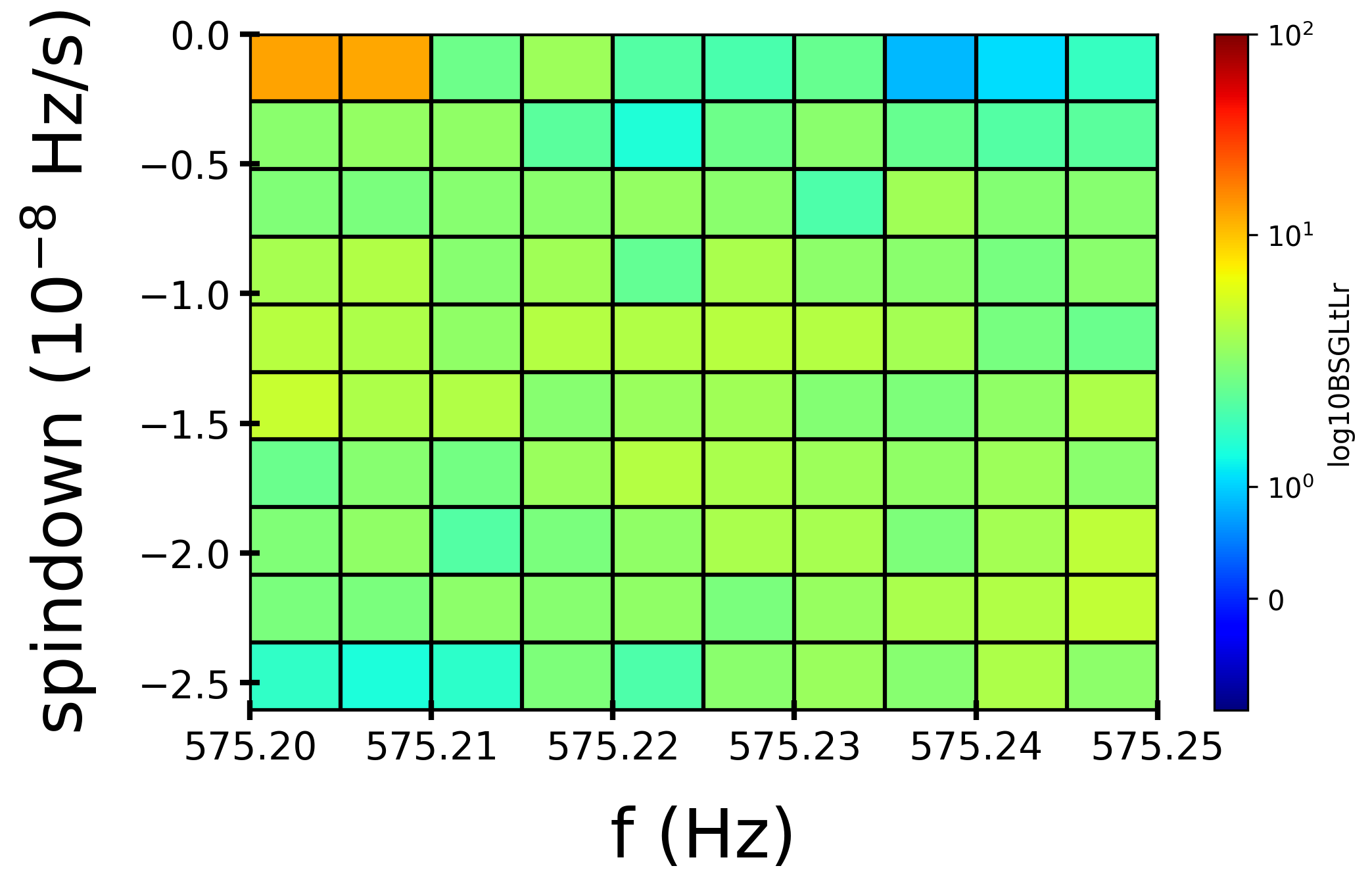}
    \caption{An example of the lowest recorded $\hat{\beta}_{\text{S/GLtLr}}$ per pixel in a slightly disturbed (top-left) band. The results shown come from the Vela Jr. search. Each cell in this figure is a ``pixel".}
    \label{fig:pixeling_example} 
\end{figure}

We take the top $n_{px}=2$  Stage-0 results from waveforms with parameters in each pixel, with $n_{px}$ determined by the  available computational budget for the follow-up.  We call these {\it{candidates}}. The lowest detection statistic in the pixels from a 50 mHz region is shown in Figure \ref{fig:pixeling_example} as an example. 

In all we select $\approx 1.5$ million candidates from the Cas A search and $\approx 5.2$ million from Vela Jr. search. Both these numbers are $< 2N_p\times N_{\textrm{bands}}$ (twice the total number of pixels) because not all pixels contribute 2 candidates -- in fact some contribute no candidate at all. This is due to the fact that the original results are themselves top-lists returned by the Einstein@Home volunteer computers, and in the presence of very large disturbances most results might be concentrated in certain frequency-spindown regions rather than being $\approx$ uniformly distributed among the pixels. 
The average value of the $\hat{\beta}_{\text{S/GLtLr}}$ detection statistic of the considered candidates increases by 4 going from 20 to 400 Hz for Cas A and increases by 2 going from 400 to 1700 Hz for Vela Jr. This is due to the a trials factor effect: we are taking the most significant 2 results from a parameter space (the pixel) that increases in size as the frequency increases.

\begin{figure}[tb]
    \includegraphics[width=1.0\columnwidth] {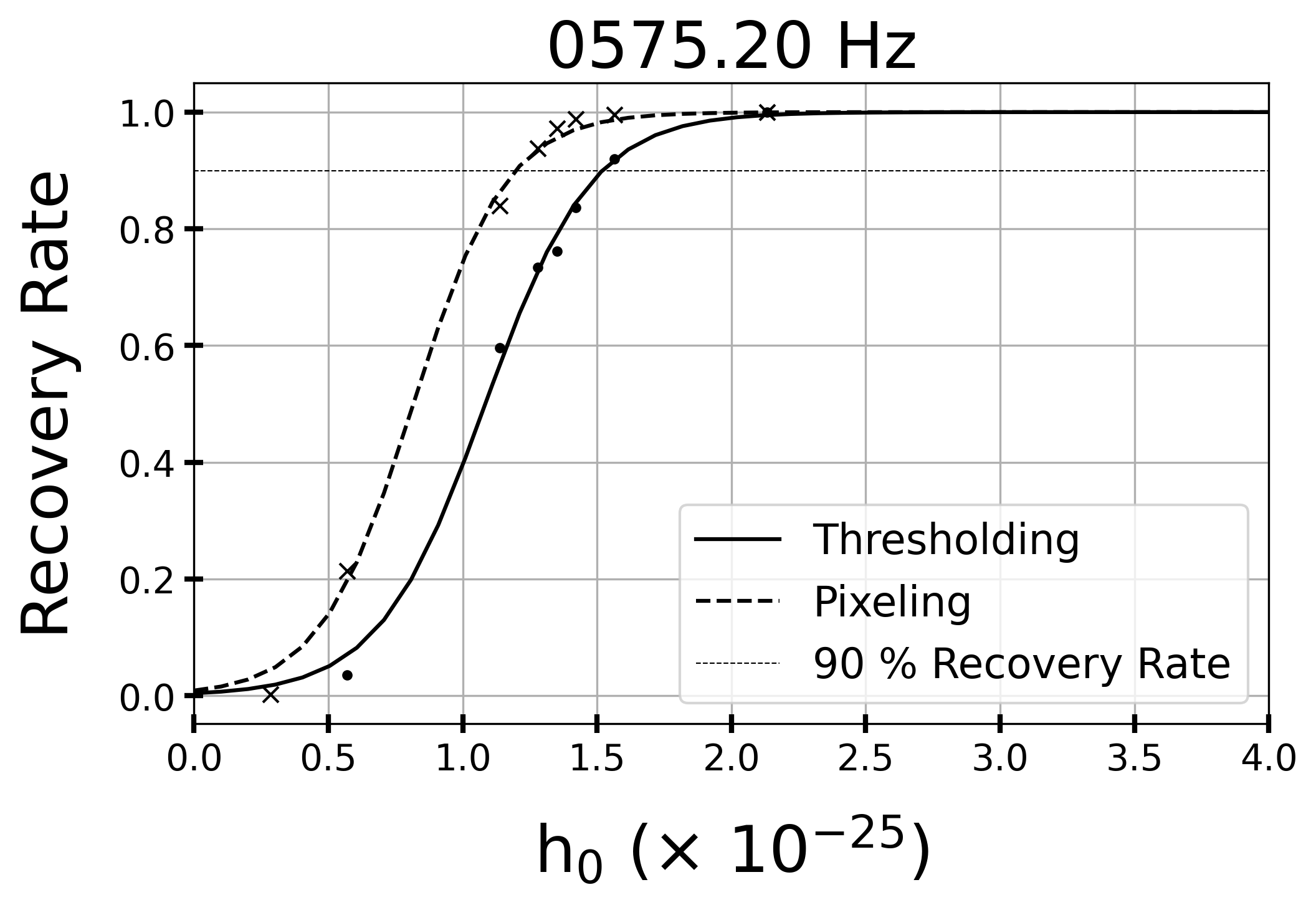}
    \caption{Comparison between the performance of the pixeling and the simple thresholding selection methods of the example slightly disturbed band.}
    \label{fig:h0_90_UL_estimate} 
\end{figure}

Compared to simple thresholding at fixed number of selected candidates, in well-behaved, nearly Gaussian noise, the pixeling procedure and simple thresholding yield comparable results.  When the data is affected by disturbances pixeling is never worse than simple thresholding and for most disturbances, which are localized in $f-\dot{f}$, the detection efficiency is enhanced with pixeling. In  Figure \ref{fig:h0_90_UL_estimate}, that features data with a typical disturbance, the 90\% detection efficiency amplitude is $\approx1.2\times10^{-25}$ using the pixeling method,  and $\approx1.6\times10^{-25}$ using the simple thresholding method. 

\begin{figure*}[]
\centering
\subfigure[Cas A]{
    \includegraphics[width=1.0\columnwidth] {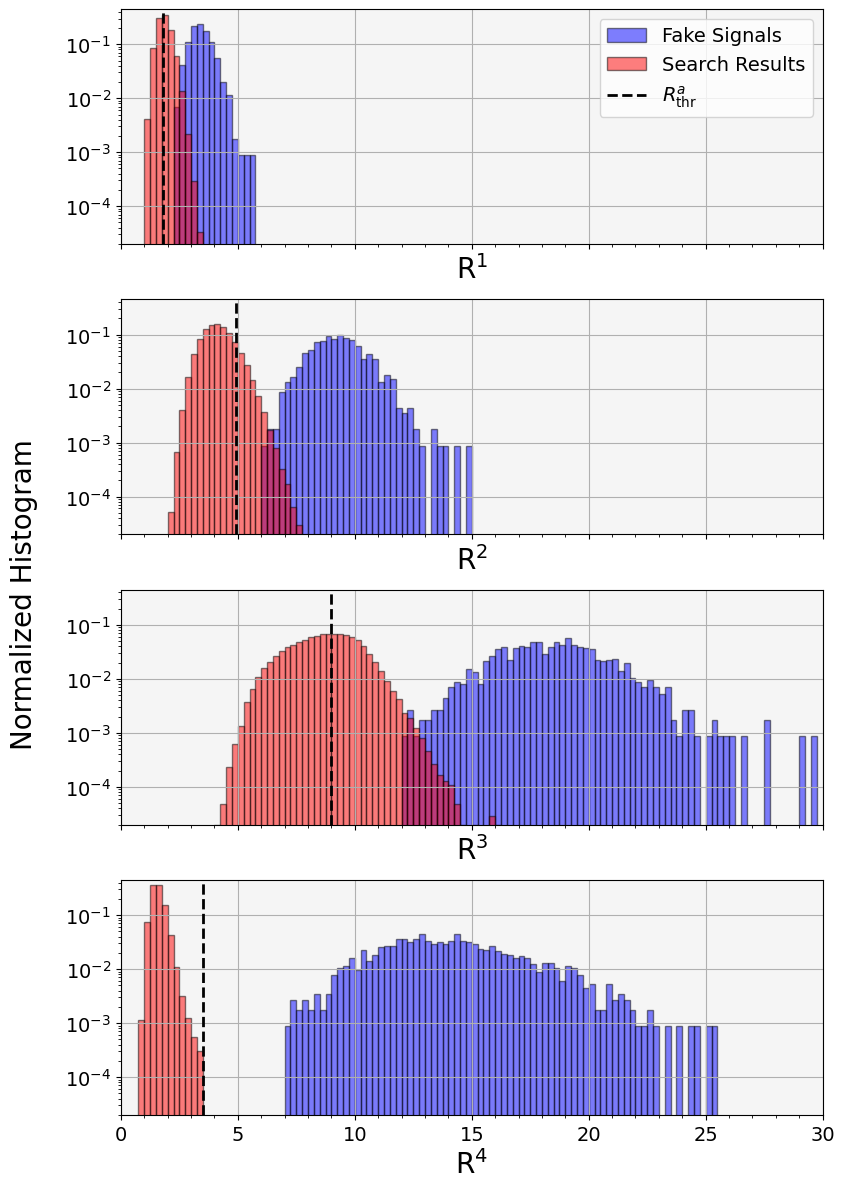}
    \label{fig:CasAR}}
    \subfigure[Vela Jr.]{
       \includegraphics[width=1.0\columnwidth] {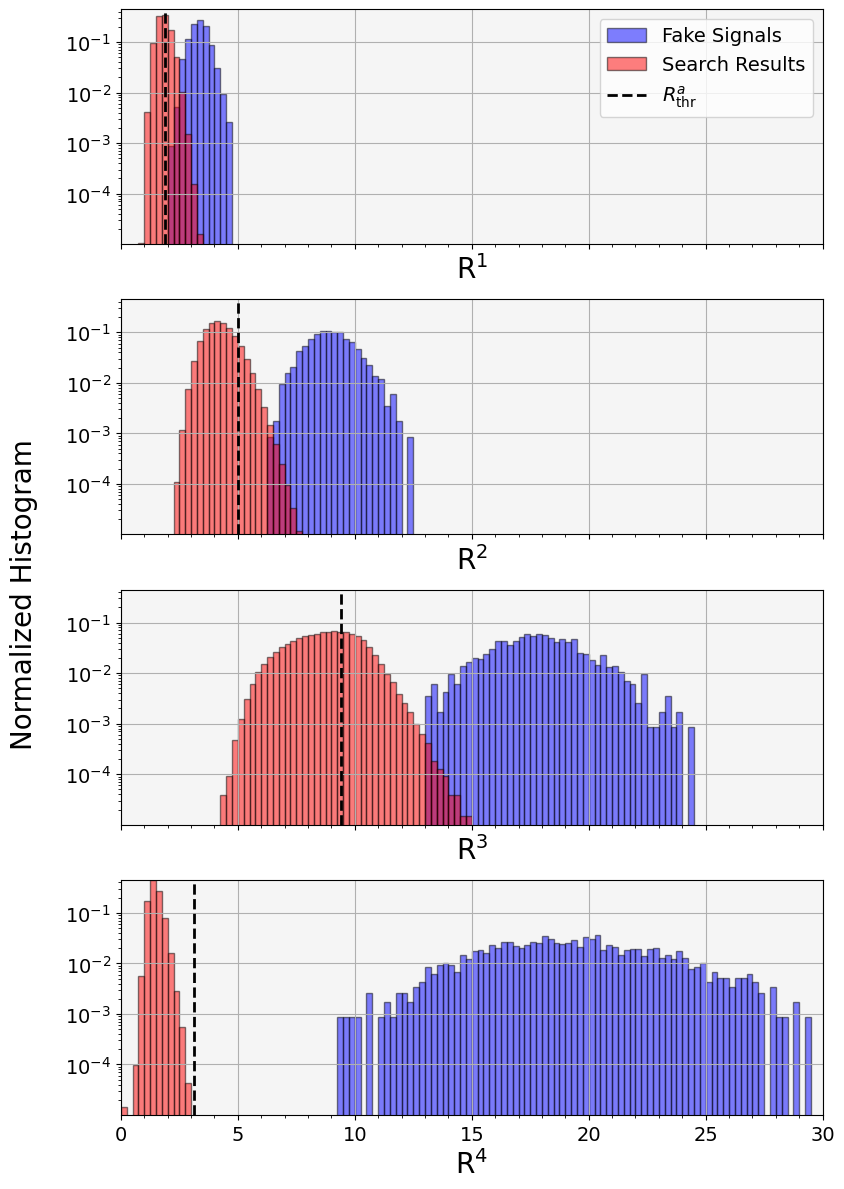}
          \label{fig:VelaR}}
    \caption{Distributions of $R^a$ of the candidates from the searches (red, left) and from the target signal population (blue, right). The vertical dashed line shows the $R^a_\mathrm{thr}$ at each stage. One can see how the noise-dominated data and the test-signal population separate more and more clearly as the hierarchical follow-ups proceed, with the data containing signals moving to higher values of $R$.}
    \label{fig:Ra_dist_casa} 
\end{figure*}
\subsection{\label{subsec:FU_stage_1} Stages 1-4}

We follow-up the candidates selected through the pixeling procedure with a cascade of three semi-coherent searches with increasing coherent time and decreasing mismatch. The surviving candidates are verified on a different data set, that data from the first half of the O3 run, O3a. The searched volume around the nominal candidate parameters at each stage is detailed in Tables \ref{tab:results_FU}. The search region for the O3a search is the containment region from the O2 search evolved to the time of the O3a data (see the details of Eq. 10 in \cite{ming2024a}). 

The search setups are determined by their effectiveness and safety in separating noise from signals and by computational feasibility. The effectiveness and safety of the thresholds on the $R^a_\mathrm{thr}$ are established based on the distributions shown in Figures  \ref{fig:Ra_dist_casa}.

The number of candidates fed to each stage and the surviving ones are detailed in Tables \ref{tab:results_FU} .

\section{\label{sec:results} Results}

\subsection{Intrinsic Gravitational Wave Amplitude Upper Limits}
\label{sec:h0ULs}

No candidate survives the last stage hence there is no evidence for a continuous-wave signal from either targets. 

We determine frequentist 90\% confidence upper limits on the intrinsic GW amplitude in every half-Hz band, $h_0^{90 \%}(f)$. $h_0^{90 \%}(f)$ is the GW amplitude such that 90\% of signals of our target population with that amplitude would have been detected by our search. To determine this value we measure the detection efficiency (or confidence) $C(h_0)$ on our target population at fixed values of $h_0$ and with a sigmoid fit we estimate the amplitude that yields 90\% detection efficiency: 
\begin{equation}
    C(h_0) = \frac{1}{1 + \text{exp}(\tfrac{\text{a}-h_0}{\text{b}})} \ .
    \label{eqn:sigmoid_vs_h0}
\end{equation}

As for previous searches \citep{Ming2019,Papa_2020midth,Ming2022,ming2024a}, MATLAB's non-linear regression function, \texttt{nlpredi},  is used to determine the optimal values for coefficients a and b, along with the corresponding covariance matrix, based on which the 95\% credible interval for the $h_0^{90\%}$ is derived. The fitting process introduces an uncertainty in  $h_0^{90 \%}$  $ \le 5\%$ for most of the half-Hz bands.
The total uncertainty in the upper limit is the sum of the fitting process  uncertainty and LIGO's calibration uncertainty, which we assume is 5 \% \citep{Cahillane2017}.

The detection efficiency is the number of recovered signals relative to the total number of signals considered. Each test signal is added to the data and then the with the test signals is treated exactly like the data from original Einstein@Home search, including data preparation, cleaning and candidate-selection with  pixeling. A signal is considered recovered if a result from the search with the test signal, and ascribable to the signal, survives 
until the last stage. In practice, since the confidence in recovering signals at our upper limit amplitudes during the follow-up stages is much higher (99.96\%) than the upper limit confidence level  ($90\%$) we neglect the false dismissal incurred by the follow-up and significantly simplify the upper limit pipeline by requiring that the test signal only survive Stage-0.

As described in Section \ref{sec:data}, in frequency bands affected by large spectral-line disturbances, we substitute the real data with Gaussian noise. For consistency we also do this after the test signal has been added to noise, hence, depending on the data removed ands on the frequency of the test signal, it might not be possible to reach the 90\% confidence level for any value of $h_0$. For frequency bands  where this happens we do not provide upper limits.  We add these bands to  the list of excluded bands given in \cite{AEIULurl}. Such bands include 6 half-Hz bands between 20 Hz and 400 Hz and 109 half-Hz bands between 400 Hz and 1700 Hz search, mostly due to the thermal excitation of vibration modes of the suspension silica fibers at 500 Hz and their harmonics around 1000 Hz and 1500 Hz \citep{Abbott2016}.

\begin{figure*}[]
\centering
\subfigure[Cas A]{
   \includegraphics[width=2 \columnwidth]{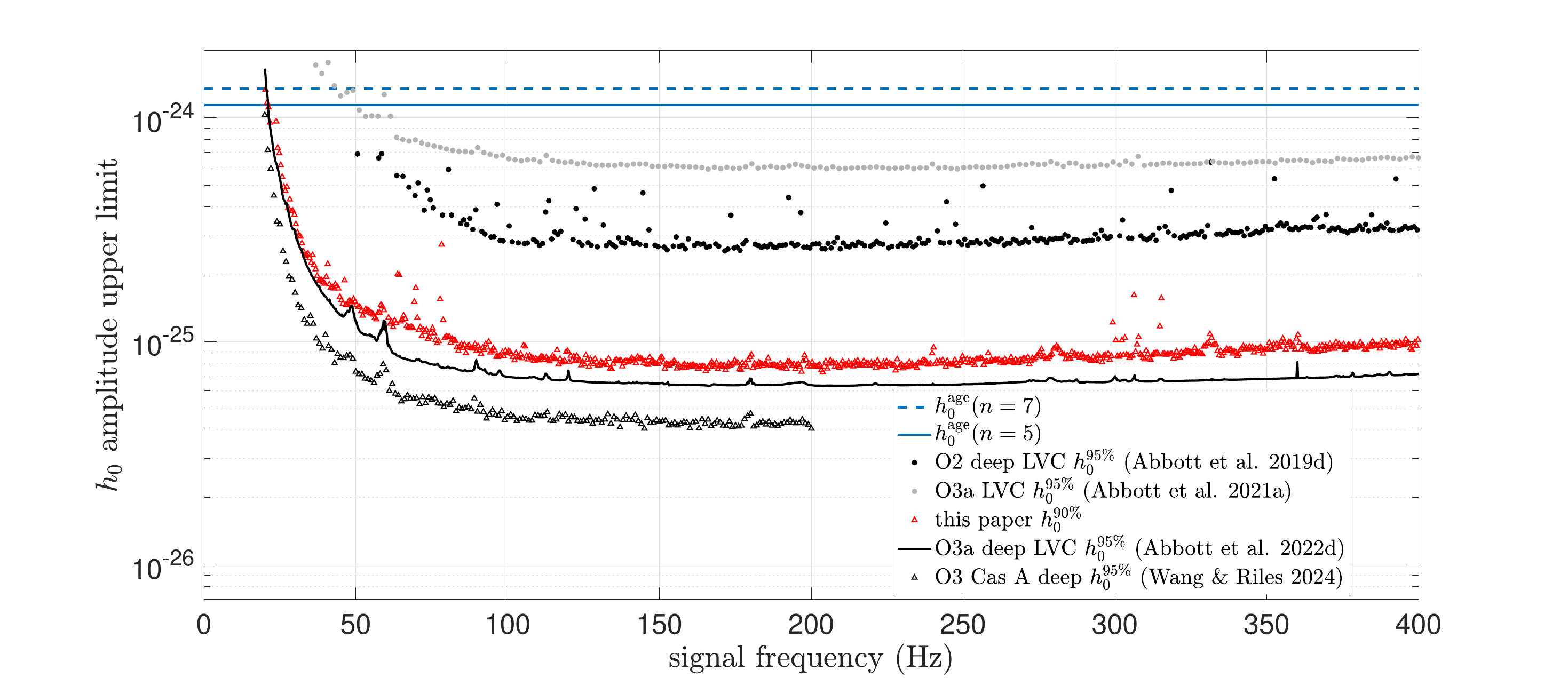}
}\\
    \subfigure[Vela Jr.]{
   \includegraphics[width=2\columnwidth]{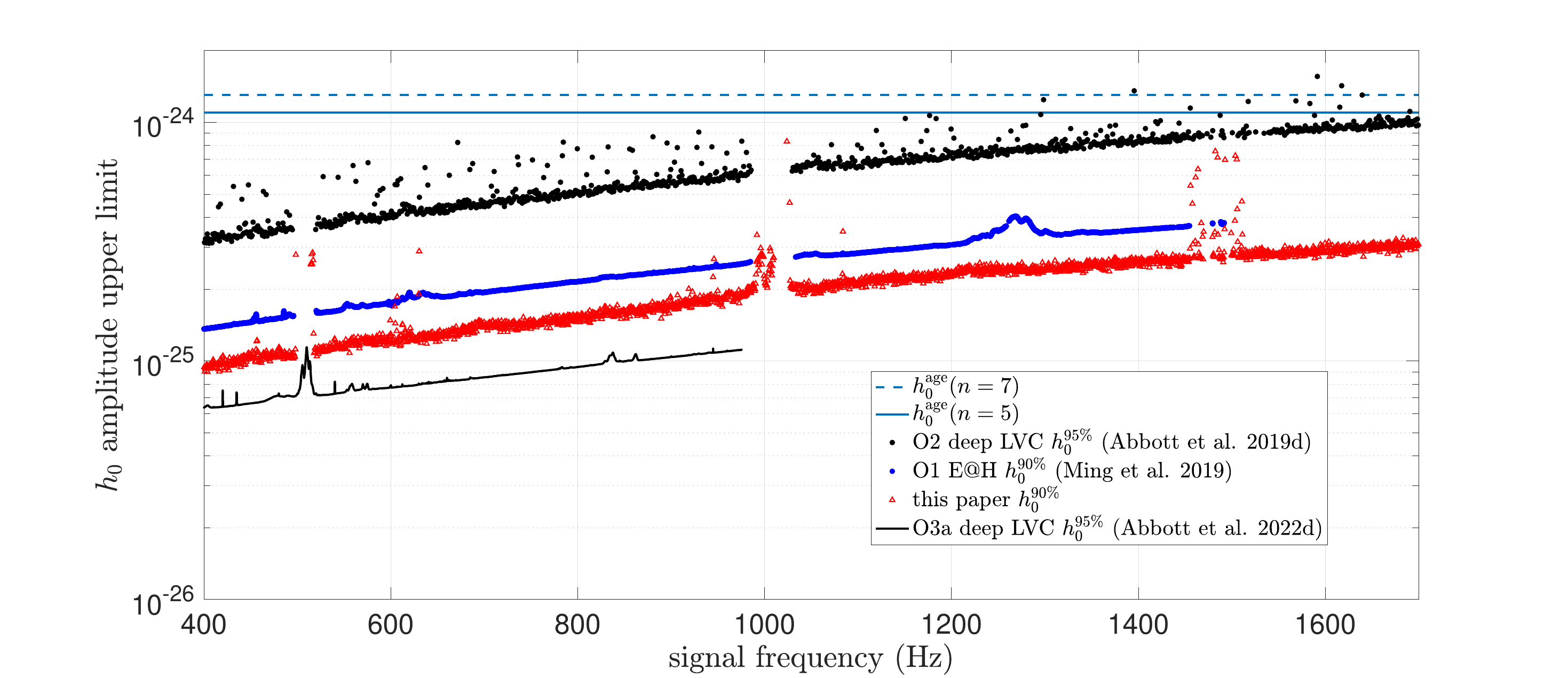}
}
\caption{90\% confidence upper limits on the gravitational wave amplitude of continuous gravitational wave signals from our searches (red triangles) as a function of frequency, compared to other recent results. The horizontal lines show the indirect age-based upper limits corresponding to braking indexes of 5 and 7. Our searches could detect signals with n as high as 7. \label{fig:h0ULs}}
\end{figure*}

The $h_0^{90\%}$ upper limits for the Cas A and Vela Jr. searches lie in  the upper $10^{-26}-$low $10^{-25}$ range across the searched frequency bands, with the most stringent upper limit being $7.3 \times 10^{-26}$ near 200 Hz for Cas A and near 400 Hz for Vela Jr. at $8.9 \times 10^{-26}$ -- see Figures \ref{fig:h0ULs}. 
Both searches are by far the most sensitive on O2 data: our Cas A search is 4 times more sensitive than the previous O2 best sensitivity achieved by \cite{LVC_O2_V}, and our Vela Jr., search is 3 times more sensitive than  \cite{LVC_O2_V}. 

Searches carried out on the more sensitive O3 data provide more stringent constraints than ours: \cite{Abbott2022_4} on Cas A and for Vela Jr. up to 976 Hz achieve an upper limit a factor $\lesssim 0.8$ smaller than ours, albeit with a factor of 1.6 more sensitive O3a. \cite{PhysRevD.110.042006} achieve the 1.6 sensitivity improvement using the larger full O3 data set, and searching a smaller 200 Hz band.

Above 976 Hz the Vela Jr. search presented here is the most sensitive search carried out to date. 

\subsection{Upper Limits Recast}
\label{sec:ULsRecast}

\begin{figure*}[]
\centering
\subfigure[Upper limits on the ellipticity]{
     \includegraphics[width=2\columnwidth]{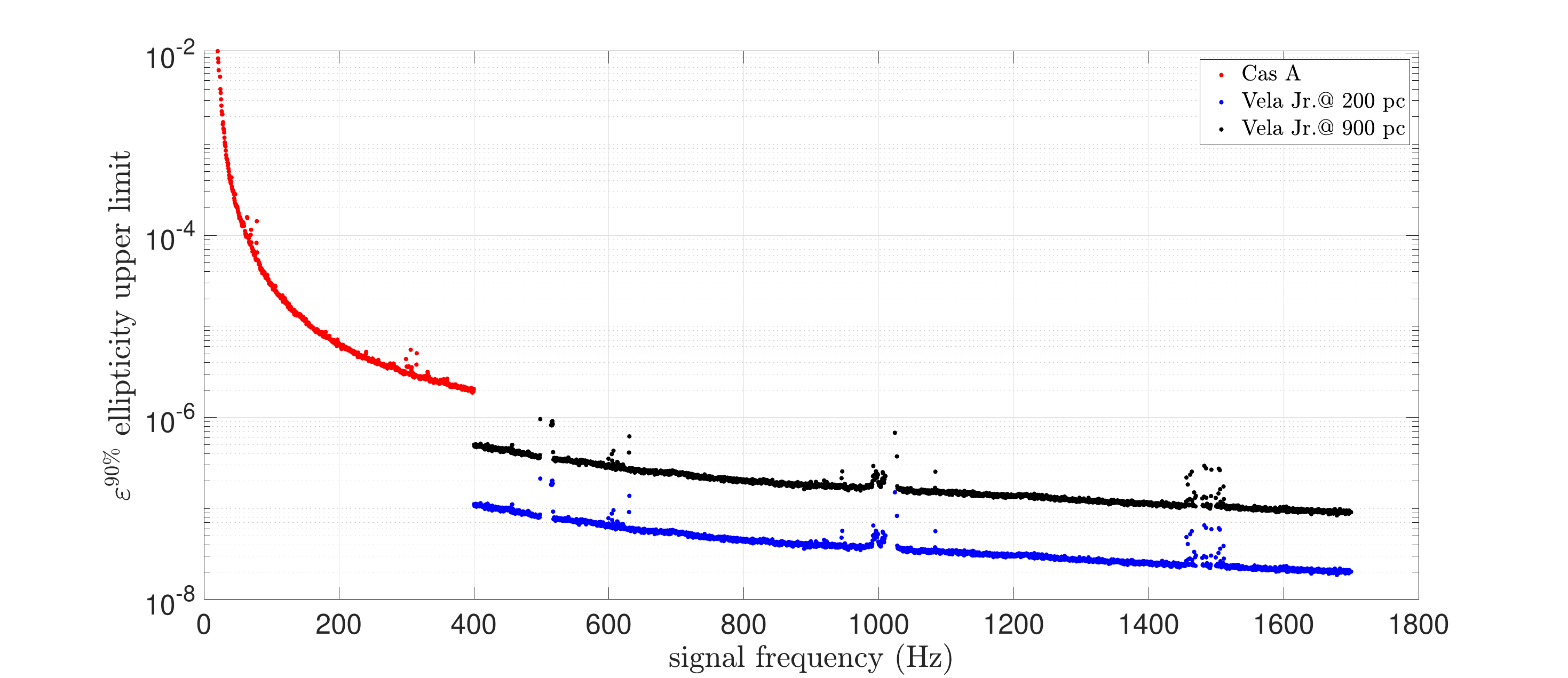}
}\\
    \subfigure[Upper limits on the r-mode amplitude]{
   \includegraphics[width=2\columnwidth]{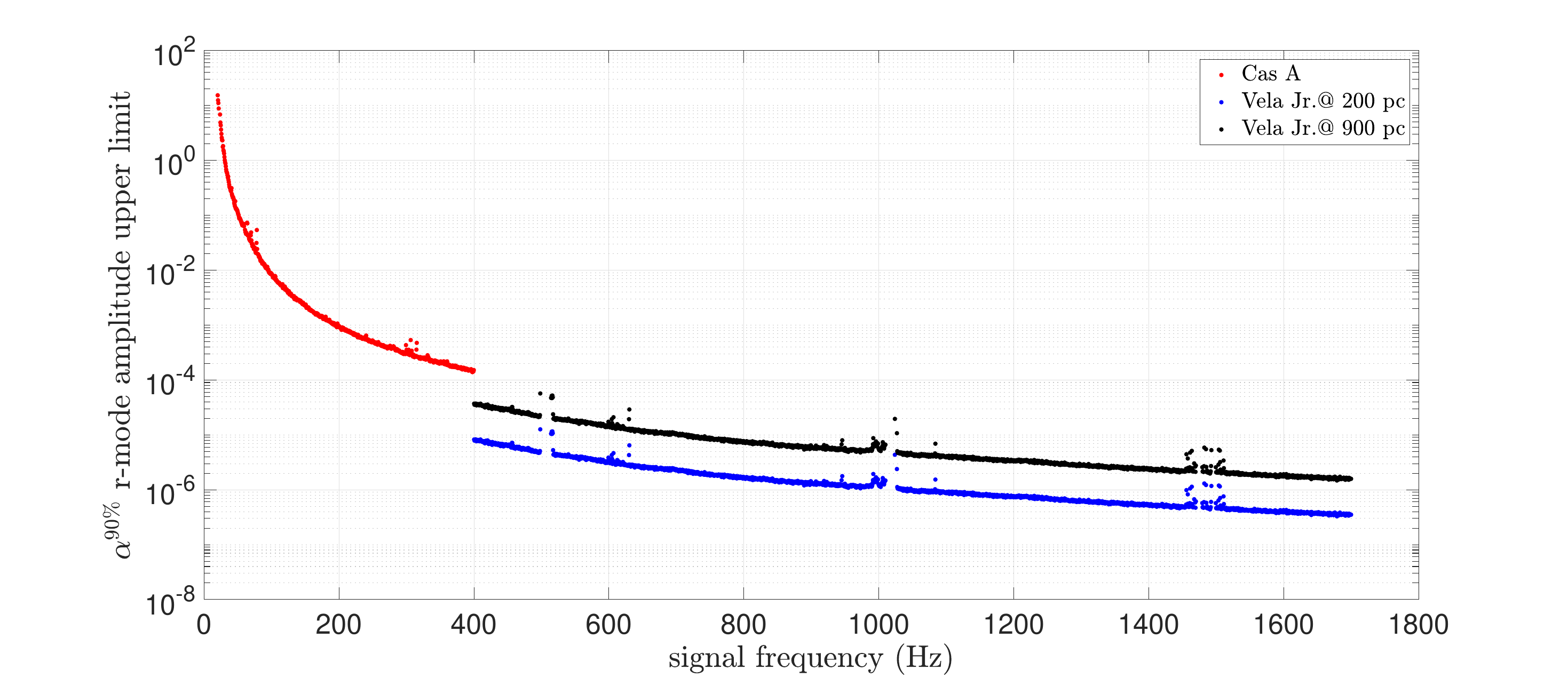}
}
\caption{90 \% upper limits on the ellipticity (a) and r-mode amplitude (b) of the two targets. For Vela Jr. we show two curves, corresponding to two distance estimates: 200 pc and 900 pc.  For Cas A, we assume 3400 pc.  \label{fig:ULsRecast}}
\end{figure*}

We can re-cast the $h_0^{90 \%}$ upper limits into equatorial ellipticity upper limits using \citep{zimmermann:1979}
\begin{equation}
    \varepsilon = \frac{c^4}{4 \pi ^2} \frac{h_0 D}{I f^2} \ ,
    \label{eqn:ellipticity_UL}
\end{equation}
where $c$ is the speed of light, $G$ is the gravitational constant, and I is the moment of inertia of the neutron star with respect to its axis of rotation. We assume the conventional value of $10^{38}$ kg m$^2$ for $I$. We use the following distances for our targets: 200 pc and 900 pc for Vela Jr. and 3.4 kpc for Cas A. The results are shown in Figure \ref{fig:ULsRecast} (a). 

Additionally, we can re-cast the $h_0^{90 \%}$  into upper limits for the r-mode amplitude \citep{Owen2010}
\begin{equation}
	\alpha = 0.028\left( \frac{h_0}{10^{-24}} \right) \left( \frac{D}{1 \  \text{kpc}} \right) \left( \frac{100 \ \text{Hz}}{f} \right)^3 .
	\label{eqn:r_mode_amp}
\end{equation}
and show them in Figure \ref{fig:ULsRecast} (b).

\subsection{Sensitivity Depth}
\label{sec:SensD}

The sensitivity depth is a useful quantity first introduced by \cite{Behnke:2014tma} to help compare different searches on data with the same noise level \citep{Dreissigacker2018}. The sensitivity depth is defined as
\begin{equation}
    \mathcal{D}^{90 \%}(f) \equiv \frac{\sqrt{S_n(f)}}{h_0^{90 \%}(f)} \ [1/\sqrt{\text{Hz}}]\ ,
    \label{eqn:sensitivity_depth}
\end{equation}
where $\sqrt{S_n(f)}$ is the noise level associated with the frequency of a putative signal. In essence, the sensitivity depth quantifies how deep the weakest detectable signal is buried into the noise floor. It depends on the search setup and on the post-processing of the results.


 The average sensitivity depth across the frequency ranges is approximately $92[{1/\sqrt{\text{Hz}}}]$ for Cas A and $89[{1/\sqrt{\text{Hz}}}]$ for Vela Jr., respectively.
Although the average mismatch of the original Einstein@Home search for the Vela Jr. is slightly smaller than for Cas A, the average sensitivity depth of search for Vela Jr. is slightly larger. 
The reason for this is that the follow-ups involve proportionally about 3.5 times more candidates for Cas A than for Vela Jr.
 
\section{\label{sec:Conclu} Conclusions}
We present results for searches for continuous gravitational waves from the neutron star in the Cas A and Vela Jr. supernova remnants. The considered waveforms cover frequencies in the range  20 - 400 Hz for Cas A and 400 - 1700Hz Vela Jr., and very broad frequency evolution scenarios including power-law frequency evolutions with braking indexes up to 7. In particular, the second-order spin-down range of our searches is $\ddot{f}\in [0,7 {f/\tau^2}]$, which is broader than the range covered by \cite{Abbott2022_4} and \cite{PhysRevD.110.042006}, who assume $[2|\dot{f}|^2/f, 7|\dot{f}|^2/f]$. Furthermore, our search range is independent of the template $\dot{f}$ value, accommodating signals that deviate from a strict power law.

The large parameter space, comprising in all  $1.3\times10^{18}$ waveforms, could be investigated thanks to the computing power donated by the volunteers of the Einstein@Home project. Promising candidates across this space were identified using a new ``pixeling'' method and followed-up in a cascade of further searches. Fewer and fewer candidates survive each stage and at the end of the fourth stage no signal candidate is left, indicating no detectable signal.  

We place upper limits on the intrinsic gravitational wave strain in half-Hz frequency bands.
Comparing with the previous most sensitive search for Cas A in O2 data \citep{LVC_O2_V}, our $h_0$ upper limit is significantly better by a factor of 4. 

Due to the great sensitivity improvement of O3 data, on parts of the parameter space, our search is less sensitive than the two searches using O3a/O3 data \citep{Abbott2022_4,PhysRevD.110.042006}. However, 
when the difference in $h_0$ sensitivity between two searches for emission from young supernova remnants is quite small (tens of percent, not orders of magnitude, like in this case), carrying out a slightly less sensitive search on different data has value: Approximately 6\% of pulsars are known to have experienced at least one glitch \citep{glitch_rev_2022}, and glitches predominantly occur in young pulsars \citep{glitch_2013}. Cas A is the youngest known supernova remnant in the Milky Way and its neutron star likely glitched since it was born roughly 330 years ago. If the Cas A neutron star glitched during the O3 run, the two O3 deep searches \cite{Abbott2022_4} and \cite{PhysRevD.110.042006} might well have missed the signal, which could have been instead picked up by this search because it did not employ a long coherence time on O3 data. 
The O3a Viterbi search for Cas A \citep{LIGOScientific:2021mwx} is based on a hidden Markov model \citep{Sun:2017zge}, which is more robust to glitches. Our search is complementary to that in that it is  an order of magnitude more sensitive, as shown by the grey dots in Figure \ref{fig:h0ULs}(a).

Below 976 Hz, the LVC O3a search for Vela Jr.  \citep{Abbott2022_4} has a better sensitivity than ours by 30 \%;  Above 976 Hz the search presented here is the most sensitive search ever performed. It  improves by 50\% on our previous deep search \citep{Ming2019} between 976 Hz and 1500 Hz, and above 1500 Hz it improves by 300\% on the best previous result which was by \cite{LVC_O2_V} also on O2 data.

We recast the upper limits on the intrinsic strain into upper limits for both the equatorial ellipticity $\varepsilon$ and the Newtonian r-mode amplitude (see Figures~\ref{fig:ULsRecast}. For Cas A, $\varepsilon^{90\%}$ is constrained below $10^{-5}$ at the frequencies higher than $\approx$ 150 Hz, dropping to $< 2\times 10^{-6}$ at 400 Hz.
For Vela Jr., assuming a distance of 200 pc, our results constrain $\varepsilon^{90\%}$ to be $< 10^{-7}$ for all investigated signal frequencies. At the highest frequency 1700 Hz, $\varepsilon^{90\%}$ is $2\times10^{-8}$. This is more than one order of magnitude lower than the physical plausible values predicted by \cite{McDanielJohnsonOwen,Gittins2021,Gittins2021_2}.

Newly-born neutron stars could present  active r-modes oscillations, leading to continuous gravitational-wave emission and fast spin-down after collapse \citep{Arras_2003,Owen:1998xg}.  
Our search for Vela Jr. covers the r-mode emitting neutron star with a  spin frequency up to $\approx 1275$ Hz, and for Cas A up to a spin frequency of $\approx$ 300 Hz. The expected r-mode amplitude $\alpha$ is quite small, with estimates ranging from $10^{-5}$ to $10^{-3}$ \citep{PhysRevD.76.064019,Haskell:2015iia}. Our results for Cas A constrain $\alpha^{90\%}$ to be smaller that $\approx 10^{-3}$ above 200 Hz reaching $\approx1\times 10^{-4}$, at 400 Hz. For Vela Jr., assuming a distance of 200 pc, our results constrain  $\alpha^{90\%}$ at 1700 Hz to be smaller than $\approx3\times 10^{-7}$, which well in the range of possible values.

\section{Acknowledgments}

We gratefully acknowledge the support of the many thousands of Einstein@Home volunteers who made this search possible. \\
We acknowledge support from the Max Planck Society for Projects QPQ10003 and QPQ10004, and the NSF grant 1816904.\\ 
A lot of post-processing is run on the ATLAS cluster at AEI Hannover. We thank Carsten Aulbert and Henning Fehrmann for their support. \\
We would like to thank the instrument-scientist and engineers of LIGO whose amazing work has produced detectors capable of probing gravitational waves so incredibly small.\\
J.A.M. acknowledges the continuing support and guidance of Charles J. Horowitz, as well as partial support from the US Department of Energy grant DE-FG02-87ER40365. \\

This research has made use of data, software and/or web tools obtained from the Gravitational Wave Open Science Center (https://www.gw-openscience.org/ ), a service of LIGO Laboratory, the LIGO Scientific Collaboration and the Virgo Collaboration. LIGO Laboratory and Advanced LIGO are funded by the United States National Science Foundation (NSF) as well as the Science and Technology Facilities Council (STFC) of the United Kingdom, the Max-Planck-Society (MPS), and the State of Niedersachsen/Germany for support of the construction of Advanced LIGO and construction and operation of the GEO600 detector. Additional support for Advanced LIGO was provided by the Australian Research Council. Virgo is funded, through the European Gravitational Observatory (EGO), by the French Centre National de Recherche Scientifique (CNRS), the Italian Istituto Nazionale di Fisica Nucleare (INFN) and the Dutch Nikhef, with contributions by institutions from Belgium, Germany, Greece, Hungary, Ireland, Japan, Monaco, Poland, Portugal, Spain. \\

\bibliography{bibliography}

\begin{thebibliography}{}
\expandafter\ifx\csname natexlab\endcsname\relax\def\natexlab#1{#1}\fi
\providecommand{\url}[1]{\href{#1}{#1}}
\providecommand{\dodoi}[1]{doi:~\href{http://doi.org/#1}{\nolinkurl{#1}}}
\providecommand{\doeprint}[1]{\href{http://ascl.net/#1}{\nolinkurl{http://ascl.net/#1}}}
\providecommand{\doarXiv}[1]{\href{https://arxiv.org/abs/#1}{\nolinkurl{https://arxiv.org/abs/#1}}}

\bibitem[{A.~G. Abac \&  other(2025)Abac \& other}]{lvk2025a}
Abac, A.~G., \& other. 2025, \bibinfo{title}{Search for continuous
  gravitational waves from known pulsars in the first part of the fourth
  LIGO-Virgo-KAGRA observing run,} \doarXiv{2501.01495}

\bibitem[{B.~P. Abbott {et~al.}(2016)Abbott {et~al.}}]{Abbott2016}
Abbott, B.~P., {et~al.} 2016, \bibinfo{title}{{Comprehensive all-sky search for
  periodic gravitational waves in the sixth science run LIGO data},} Phys. Rev.
  D, 94, 042002, \dodoi{10.1103/PhysRevD.94.042002}

\bibitem[{B.~P. Abbott {et~al.}(2019{\natexlab{a}})Abbott
  {et~al.}}]{Abbott2019}
Abbott, B.~P., {et~al.} 2019{\natexlab{a}}, \bibinfo{title}{{Narrow-band search
  for gravitational waves from known pulsars using the second LIGO observing
  run},} Phys. Rev. D, 99, 122002, \dodoi{10.1103/PhysRevD.99.122002}

\bibitem[{B.~P. Abbott {et~al.}(2019{\natexlab{b}})Abbott
  {et~al.}}]{Abbott2019_2}
Abbott, B.~P., {et~al.} 2019{\natexlab{b}}, \bibinfo{title}{{Searches for
  Gravitational Waves from Known Pulsars at Two Harmonics in 2015–2017 LIGO
  Data},} ApJ, 879, 10, \dodoi{10.3847/1538-4357/ab20cb}

\bibitem[{B.~P. Abbott {et~al.}(2019{\natexlab{c}})Abbott
  {et~al.}}]{Abbott2019_3}
Abbott, B.~P., {et~al.} 2019{\natexlab{c}}, \bibinfo{title}{{Search for
  gravitational waves from Scorpius X-1 in the second Advanced LIGO observing
  run with an improved hidden Markov model},} Phys. Rev. D, 100, 122002,
  \dodoi{10.1103/PhysRevD.100.122002}

\bibitem[{B.~P. Abbott {et~al.}(2019{\natexlab{d}})Abbott {et~al.}}]{LVC_O2_V}
Abbott, B.~P., {et~al.} 2019{\natexlab{d}}, \bibinfo{title}{Searches for
  Continuous Gravitational Waves from 15 Supernova Remnants and Fomalhaut b
  with Advanced LIGO*,} The Astrophysical Journal, 875, 122,
  \dodoi{10.3847/1538-4357/ab113b}

\bibitem[{R. Abbott {et~al.}(2021{\natexlab{a}})Abbott
  {et~al.}}]{LIGOScientific:2021mwx}
Abbott, R., {et~al.} 2021{\natexlab{a}}, \bibinfo{title}{Searches for
  Continuous Gravitational Waves from Young Supernova Remnants in the Early
  Third Observing Run of Advanced LIGO and Virgo,} The Astrophysical Journal,
  921, 80, \dodoi{10.3847/1538-4357/ac17ea}

\bibitem[{R. Abbott {et~al.}(2021{\natexlab{b}})Abbott
  {et~al.}}]{Abbott:2021boh}
Abbott, R., {et~al.} 2021{\natexlab{b}}, \bibinfo{title}{{Open data from the
  first and second observing runs of Advanced LIGO and Advanced Virgo},}
  SoftwareX, 13, 012021, \dodoi{https://doi.org/10.1016/j.softx.2021.100658}

\bibitem[{R. Abbott {et~al.}(2022{\natexlab{a}})Abbott {et~al.}}]{Abbott2022}
Abbott, R., {et~al.} 2022{\natexlab{a}}, \bibinfo{title}{Narrowband Searches
  for Continuous and Long-duration Transient Gravitational Waves from Known
  Pulsars in the LIGO-Virgo Third Observing Run,} The Astrophysical Journal,
  932, 133, \dodoi{10.3847/1538-4357/ac6ad0}

\bibitem[{R. Abbott {et~al.}(2022{\natexlab{b}})Abbott {et~al.}}]{Abbott2022_2}
Abbott, R., {et~al.} 2022{\natexlab{b}}, \bibinfo{title}{{Search for Continuous
  Gravitational Waves from 20 accreting Millisecond X-ray Pulsars in O3 LIGO
  data},} Phys. Rev. D, 105, 022002, \dodoi{10.1103/PhysRevD.105.022002}

\bibitem[{R. Abbott {et~al.}(2022{\natexlab{c}})Abbott {et~al.}}]{Abbott2022_3}
Abbott, R., {et~al.} 2022{\natexlab{c}}, \bibinfo{title}{All-sky search for
  continuous gravitational waves from isolated neutron stars using Advanced
  LIGO and Advanced Virgo O3 data,} Phys. Rev. D, 106, 102008,
  \dodoi{10.1103/PhysRevD.106.102008}

\bibitem[{R. Abbott {et~al.}(2022{\natexlab{d}})Abbott {et~al.}}]{Abbott2022_4}
Abbott, R., {et~al.} 2022{\natexlab{d}}, \bibinfo{title}{{Search of the early
  O3 LIGO data for continuous gravitational waves from the Cassiopeia A and
  Vela Jr. supernova remnants},} Phys. Rev. D, 105, 082005,
  \dodoi{10.1103/PhysRevD.105.082005}

\bibitem[{R. Abbott {et~al.}(2022{\natexlab{e}})Abbott
  {et~al.}}]{KAGRA:2022osp}
Abbott, R., {et~al.} 2022{\natexlab{e}}, \bibinfo{title}{{Search for continuous
  gravitational wave emission from the Milky~Way center in O3 LIGO-Virgo
  data},} Phys. Rev. D, 106, 042003, \dodoi{10.1103/PhysRevD.106.042003}

\bibitem[{R. Abbott {et~al.}(2022{\natexlab{f}})Abbott
  {et~al.}}]{LIGOScientific:2022enz}
Abbott, R., {et~al.} 2022{\natexlab{f}}, \bibinfo{title}{{Model-based
  Cross-correlation Search for Gravitational Waves from the Low-mass X-Ray
  Binary Scorpius X-1 in LIGO O3 Data},} Astrophys. J. Lett., 941, L30,
  \dodoi{10.3847/2041-8213/aca1b0}

\bibitem[{R. Abbott {et~al.}(2022{\natexlab{g}})Abbott
  {et~al.}}]{LIGOScientific:2021rnv}
Abbott, R., {et~al.} 2022{\natexlab{g}}, \bibinfo{title}{{All-sky search for
  gravitational wave emission from scalar boson clouds around spinning black
  holes in LIGO O3 data},} Phys. Rev. D, 105, 102001,
  \dodoi{10.1103/PhysRevD.105.102001}

\bibitem[{R. Abbott {et~al.}(2023)Abbott {et~al.}}]{Abbott_2023}
Abbott, R., {et~al.} 2023, \bibinfo{title}{Open Data from the Third Observing
  Run of LIGO, Virgo, KAGRA, and GEO,} The Astrophysical Journal Supplement
  Series, 267, 29, \dodoi{10.3847/1538-4365/acdc9f}

\bibitem[{ AEI(2023)AEI}]{Atlas}
AEI. 2023, \bibinfo{title}{Computing and ATLAS,},
  \url{https://www.aei.mpg.de/25950/computer-clusters}

\bibitem[{G.~E. Allen {et~al.}(2014)Allen, Chow, DeLaney, Filipovi?, Houck,
  Pannuti, \& Stage}]{Allen2015}
Allen, G.~E., Chow, K., DeLaney, T., {et~al.} 2014, \bibinfo{title}{ON THE
  EXPANSION RATE, AGE, AND DISTANCE OF THE SUPERNOVA REMNANT G266.2?1.2 (Vela
  Jr.),} The Astrophysical Journal, 798, 82, \dodoi{10.1088/0004-637x/798/2/82}

\bibitem[{D. Anderson(2004)Anderson}]{Anderson2004}
Anderson, D. 2004, in Fifth IEEE/ACM International Workshop on Grid Computing,
  4--10, \dodoi{10.1109/GRID.2004.14}

\bibitem[{D.~P. Anderson {et~al.}(2006)Anderson, Christensen, \&
  Allen}]{Anderson2006}
Anderson, D.~P., Christensen, C., \& Allen, B. 2006, in Proceedings of the 2006
  ACM/IEEE Conference on Supercomputing, SC '06 (New York, NY, USA: Association
  for Computing Machinery), 126–es, \dodoi{10.1145/1188455.1188586}

\bibitem[{N. Andersson {et~al.}(1999)Andersson, Kokkotas, \&
  Stergioulas}]{Andersson1999}
Andersson, N., Kokkotas, K.~D., \& Stergioulas, N. 1999, \bibinfo{title}{On the
  Relevance of the r-Mode Instability for Accreting Neutron Stars and White
  Dwarfs,} The Astrophysical Journal, 516, 307, \dodoi{10.1086/307082}

\bibitem[{P. Arras {et~al.}(2003)Arras, Flanagan, Morsink, Schenk, Teukolsky,
  \& Wasserman}]{Arras_2003}
Arras, P., Flanagan, E.~E., Morsink, S.~M., {et~al.} 2003,
  \bibinfo{title}{Saturation of the r-Mode Instability,} The Astrophysical
  Journal, 591, 1129, \dodoi{10.1086/374657}

\bibitem[{A. Arvanitaki {et~al.}(2015)Arvanitaki, Huang, \&
  Van~Tilburg}]{Arvanitaki2015}
Arvanitaki, A., Huang, J., \& Van~Tilburg, K. 2015, \bibinfo{title}{Searching
  for dilaton dark matter with atomic clocks,} Phys. Rev. D, 91, 015015,
  \dodoi{10.1103/PhysRevD.91.015015}

\bibitem[{A. Ashok {et~al.}(2024)Ashok, Covas, Prix, \& Papa}]{Ashok2024}
Ashok, A., Covas, P.~B., Prix, R., \& Papa, M.~A. 2024,
  \bibinfo{title}{Bayesian $\mathcal{F}$-statistic-based parameter estimation
  of continuous gravitational waves from known pulsars,} Phys. Rev. D, 109,
  104002, \dodoi{10.1103/PhysRevD.109.104002}

\bibitem[{A. Ashok {et~al.}(2021)Ashok, Beheshtipour, Papa, Freire, Steltner,
  Machenschalk, Behnke, Allen, \& Prix}]{Ashok2021}
Ashok, A., Beheshtipour, B., Papa, M.~A., {et~al.} 2021, \bibinfo{title}{New
  Searches for Continuous Gravitational Waves from Seven Fast Pulsars,} The
  Astrophysical Journal, 923, 85, \dodoi{10.3847/1538-4357/ac2582}

\bibitem[{B. Behnke {et~al.}(2015)Behnke, Papa, \& Prix}]{Behnke:2014tma}
Behnke, B., Papa, M.~A., \& Prix, R. 2015, \bibinfo{title}{{Postprocessing
  methods used in the search for continuous gravitational-wave signals from the
  Galactic Center},} Phys. Rev. D, 91, 064007,
  \dodoi{10.1103/PhysRevD.91.064007}

\bibitem[{R. Bondarescu {et~al.}(2007)Bondarescu, Teukolsky, \&
  Wasserman}]{PhysRevD.76.064019}
Bondarescu, R., Teukolsky, S.~A., \& Wasserman, I. 2007, \bibinfo{title}{Spin
  evolution of accreting neutron stars: Nonlinear development of the $r$-mode
  instability,} Phys. Rev. D, 76, 064019, \dodoi{10.1103/PhysRevD.76.064019}

\bibitem[{P.~R. {Brady} \& T. {Creighton}(2000){Brady} \&
  {Creighton}}]{2000PhRvD..61h2001B}
{Brady}, P.~R., \& {Creighton}, T. 2000, \bibinfo{title}{{Searching for
  periodic sources with LIGO. II. Hierarchical searches},} \prd, 61, 082001,
  \dodoi{10.1103/PhysRevD.61.082001}

\bibitem[{P.~R. Brady {et~al.}(1998)Brady, Creighton, Cutler, \&
  Schutz}]{Brady:1997ji}
Brady, P.~R., Creighton, T., Cutler, C., \& Schutz, B.~F. 1998,
  \bibinfo{title}{{Searching for periodic sources with LIGO},} Phys. Rev. D,
  57, 2101, \dodoi{10.1103/PhysRevD.57.2101}

\bibitem[{E.~F. Brown \& G. Ushomirsky(2000)Brown \& Ushomirsky}]{Brown2000}
Brown, E.~F., \& Ushomirsky, G. 2000, \bibinfo{title}{Constraints on the Steady
  State r-Mode Amplitude in Neutron Star Transients,} The Astrophysical
  Journal, 536, 915, \dodoi{10.1086/308969}

\bibitem[{C. Cahillane {et~al.}(2017)Cahillane, Betzwieser, Brown, Goetz, Hall,
  Izumi, Kandhasamy, Karki, Kissel, Mendell, Savage, Tuyenbayev, Urban, Viets,
  Wade, \& Weinstein}]{Cahillane2017}
Cahillane, C., Betzwieser, J., Brown, D.~A., {et~al.} 2017,
  \bibinfo{title}{Calibration uncertainty for Advanced LIGO's first and second
  observing runs,} Phys. Rev. D, 96, 102001, \dodoi{10.1103/PhysRevD.96.102001}

\bibitem[{C.~J. Clark {et~al.}(2023)Clark {et~al.}}]{Clark:2022tjf}
Clark, C.~J., {et~al.} 2023, \bibinfo{title}{{The TRAPUM L-band survey for
  pulsars in Fermi-LAT gamma-ray sources},} Mon. Not. Roy. Astron. Soc., 519,
  5590, \dodoi{10.1093/mnras/stac3742}

\bibitem[{P.~B. Covas {et~al.}(2024)Covas, Papa, \& Prix}]{covas2024}
Covas, P.~B., Papa, M.~A., \& Prix, R. 2024, \bibinfo{title}{Search for
  continuous gravitational waves from unknown neutron stars in binary systems
  with long orbital periods in O3 data,} \doarXiv{2409.16196}

\bibitem[{P.~B. Covas {et~al.}(2022)Covas, Papa, Prix, \& Owen}]{Covas2022}
Covas, P.~B., Papa, M.~A., Prix, R., \& Owen, B.~J. 2022,
  \bibinfo{title}{Constraints on r-modes and Mountains on Millisecond Neutron
  Stars in Binary Systems,} The Astrophysical Journal Letters, 929, L19,
  \dodoi{10.3847/2041-8213/ac62d7}

\bibitem[{V. Dergachev \& M. Papa(2020)Dergachev \& Papa}]{Dergachev2020}
Dergachev, V., \& Papa, M. 2020, \bibinfo{title}{{Results from the First
  All-Sky Search for Continuous Gravitational Waves from Small-Ellipticity
  Sources},} Phys. Rev. Lett., 125, 171101,
  \dodoi{10.1103/PhysRevLett.125.171101}

\bibitem[{V. Dergachev \& M.~A. Papa(2021)Dergachev \& Papa}]{Dergachev2021}
Dergachev, V., \& Papa, M.~A. 2021, \bibinfo{title}{Search for continuous
  gravitational waves from small-ellipticity sources at low frequencies,} Phys.
  Rev. D, 104, 043003, \dodoi{10.1103/PhysRevD.104.043003}

\bibitem[{V. Dergachev \& M.~A. Papa(2023)Dergachev \& Papa}]{Dergachev2023}
Dergachev, V., \& Papa, M.~A. 2023, \bibinfo{title}{{Frequency-Resolved Atlas
  of the Sky in Continuous Gravitational Waves},} Phys. Rev. X, 13, 021020,
  \dodoi{10.1103/PhysRevX.13.021020}

\bibitem[{V. Dergachev {et~al.}(2019)Dergachev, Papa, Steltner, \&
  Eggenstein}]{Dergachev2019}
Dergachev, V., Papa, M.~A., Steltner, B., \& Eggenstein, H.-B. 2019,
  \bibinfo{title}{Loosely coherent search in LIGO O1 data for continuous
  gravitational waves from Terzan 5 and the Galactic Center,} Phys. Rev. D, 99,
  084048, \dodoi{10.1103/PhysRevD.99.084048}

\bibitem[{C. Dreissigacker {et~al.}(2018)Dreissigacker, Prix, \&
  Wette}]{Dreissigacker2018}
Dreissigacker, C., Prix, R., \& Wette, K. 2018, \bibinfo{title}{Fast and
  accurate sensitivity estimation for continuous-gravitational-wave searches,}
  Phys. Rev. D, 98, 084058, \dodoi{10.1103/PhysRevD.98.084058}

\bibitem[{ Einstein@Home(2023)Einstein@Home}]{E@H}
Einstein@Home. 2023, \bibinfo{title}{Einstein@Home,},
  \url{https://einsteinathome.org/}

\bibitem[{R.~A. Fesen {et~al.}(2006)Fesen, Hammell, Morse, Chevalier,
  Borkowski, Dopita, Gerardy, Lawrence, Raymond, \& van~den Bergh}]{Fesen2006}
Fesen, R.~A., Hammell, M.~C., Morse, J., {et~al.} 2006, \bibinfo{title}{The
  Expansion Asymmetry and Age of the Cassiopeia A Supernova Remnant*,} The
  Astrophysical Journal, 645, 283, \dodoi{10.1086/504254}

\bibitem[{F. Gittins \& N. Andersson(2021)Gittins \& Andersson}]{Gittins2021_2}
Gittins, F., \& Andersson, N. 2021, \bibinfo{title}{{Modelling neutron star
  mountains in relativity},} Monthly Notices of the Royal Astronomical Society,
  507, 116, \dodoi{10.1093/mnras/stab2048}

\bibitem[{F. Gittins \& N. Andersson(2023)Gittins \& Andersson}]{Gittins2023}
Gittins, F., \& Andersson, N. 2023, \bibinfo{title}{{The r-modes of slowly
  rotating, stratified neutron stars},} Monthly Notices of the Royal
  Astronomical Society, 521, 3043, \dodoi{10.1093/mnras/stad672}

\bibitem[{F. Gittins {et~al.}(2021)Gittins, Andersson, \& Jones}]{Gittins2021}
Gittins, F., Andersson, N., \& Jones, D. 2021, \bibinfo{title}{{Modelling
  neutron star mountains},} Mon. Notices Royal Astron. Soc., 500, 5570,
  \dodoi{10.1093/mnras/staa3635}

\bibitem[{B. Haskell(2015)Haskell}]{Haskell:2015iia}
Haskell, B. 2015, \bibinfo{title}{{R-modes in neutron stars: Theory and
  observations},} Int. J. Mod. Phys. E, 24, 1541007,
  \dodoi{10.1142/S0218301315410074}

\bibitem[{B. Haskell {et~al.}(2014)Haskell, Glampedakis, \&
  Andersson}]{Haskell2014}
Haskell, B., Glampedakis, K., \& Andersson, N. 2014, \bibinfo{title}{{A new
  mechanism for saturating unstable r modes in neutron stars},} Monthly Notices
  of the Royal Astronomical Society, 441, 1662, \dodoi{10.1093/mnras/stu535}

\bibitem[{B. Haskell {et~al.}(2006)Haskell, Jones, \& Andersson}]{Haskell2006}
Haskell, B., Jones, D.~I., \& Andersson, N. 2006, \bibinfo{title}{{Mountains on
  neutron stars: accreted versus non-accreted crusts},} Monthly Notices of the
  Royal Astronomical Society, 373, 1423,
  \dodoi{10.1111/j.1365-2966.2006.10998.x}

\bibitem[{W. Ho \& C. Heinke(2009)Ho \& Heinke}]{Ho2009}
Ho, W., \& Heinke, C. 2009, \bibinfo{title}{A neutron star with a carbon
  atmosphere in the Cassiopeia A supernova remnant,} Nature, 462, 71,
  \dodoi{10.1038/nature08525}

\bibitem[{C. Horowitz {et~al.}(2020)Horowitz, Papa, \& Reddy}]{Horowitz2020}
Horowitz, C., Papa, M., \& Reddy, S. 2020, \bibinfo{title}{Search for compact
  dark matter objects in the solar system with LIGO data,} Physics Letters B,
  800, 135072, \dodoi{https://doi.org/10.1016/j.physletb.2019.135072}

\bibitem[{C.~J. Horowitz \& S. Reddy(2019)Horowitz \& Reddy}]{Horowitz2019}
Horowitz, C.~J., \& Reddy, S. 2019, \bibinfo{title}{Gravitational Waves from
  Compact Dark Objects in Neutron Stars,} Phys. Rev. Lett., 122, 071102,
  \dodoi{10.1103/PhysRevLett.122.071102}

\bibitem[{T.~J. Hutchins \& D.~I. Jones(2023)Hutchins \& Jones}]{Hutchins2022}
Hutchins, T.~J., \& Jones, D.~I. 2023, \bibinfo{title}{{Gravitational radiation
  from thermal mountains on accreting neutron stars: sources of temperature
  non-axisymmetry},} Monthly Notices of the Royal Astronomical Society, 522,
  226, \dodoi{10.1093/mnras/stad967}

\bibitem[{A.~F. Iyudin {et~al.}(1998)Iyudin, Schoenfelder, Bennett, Bloemen,
  Diehl, Hermsen, Lichti, van~der Meulen, Ryan, \& Winkler}]{Iyudin1998}
Iyudin, A.~F., Schoenfelder, V., Bennett, K., {et~al.} 1998,
  \bibinfo{title}{{Emission from 44Ti associated with a previously unknown
  Galactic supernova},} Nature, 396, 142, \dodoi{10.1038/24106}

\bibitem[{P. Jaranowski {et~al.}(1998)Jaranowski, Kr\'olak, \&
  Schutz}]{JKS1998}
Jaranowski, P., Kr\'olak, A., \& Schutz, B.~F. 1998, \bibinfo{title}{Data
  analysis of gravitational-wave signals from spinning neutron stars: The
  signal and its detection,} Phys. Rev. D, 58, 063001,
  \dodoi{10.1103/PhysRevD.58.063001}

\bibitem[{N. Johnson-McDaniel \& B. Owen(2013)Johnson-McDaniel \&
  Owen}]{McDanielJohnsonOwen}
Johnson-McDaniel, N., \& Owen, B. 2013, \bibinfo{title}{{Maximum elastic
  deformations of relativistic stars},} Phys. Rev. D, 88, 044004

\bibitem[{D. Keitel(2016)Keitel}]{Keitel:2015ova}
Keitel, D. 2016, \bibinfo{title}{{Robust semicoherent searches for continuous
  gravitational waves with noise and signal models including hours to days long
  transients},} Phys. Rev., D93, 084024, \dodoi{10.1103/PhysRevD.93.084024}

\bibitem[{D. Keitel {et~al.}(2014)Keitel, Prix, Papa, Leaci, \&
  Siddiqi}]{Keitel:2013wga}
Keitel, D., Prix, R., Papa, M.~A., Leaci, P., \& Siddiqi, M. 2014,
  \bibinfo{title}{{Search for continuous gravitational waves: Improving
  robustness versus instrumental artifacts},}
  \dodoi{10.1103/PhysRevD.89.064023}

\bibitem[{L. Lindblom \& B.~J. Owen(2020)Lindblom \& Owen}]{Lindblom2020}
Lindblom, L., \& Owen, B.~J. 2020, \bibinfo{title}{Directed searches for
  continuous gravitational waves from twelve supernova remnants in data from
  Advanced LIGO's second observing run,} Phys. Rev. D, 101, 083023,
  \dodoi{10.1103/PhysRevD.101.083023}

\bibitem[{ {Mignani, R. P.} {et~al.}(2007){Mignani, R. P.}, {De Luca, A.},
  {Zaggia, S.}, {Sester, D.}, {Pellizzoni, A.}, {Mereghetti, S.}, \& {Caraveo,
  P. A.}}]{Mignani2007}
{Mignani, R. P.}, {De Luca, A.}, {Zaggia, S.}, {et~al.} 2007,
  \bibinfo{title}{VLT observations of the central compact object in the Vela
  Jr. supernova remnant *,} Astronomy and Astrophysics, 473, 883,
  \dodoi{10.1051/0004-6361:20077768}

\bibitem[{A.~L. Miller {et~al.}(2024)Miller, Aggarwal, Clesse, De~Lillo,
  Sachdev, Astone, Palomba, Piccinni, \& Pierini}]{Miller:2024fpo}
Miller, A.~L., Aggarwal, N., Clesse, S., {et~al.} 2024,
  \bibinfo{title}{{Gravitational Wave Constraints on Planetary-Mass Primordial
  Black Holes Using LIGO O3a Data},} Phys. Rev. Lett., 133, 111401,
  \dodoi{10.1103/PhysRevLett.133.111401}

\bibitem[{M. Millhouse {et~al.}(2020)Millhouse, Strang, \&
  Melatos}]{Millhouse2020}
Millhouse, M., Strang, L., \& Melatos, A. 2020, \bibinfo{title}{Search for
  gravitational waves from 12 young supernova remnants with a hidden Markov
  model in Advanced LIGO's second observing run,} Phys. Rev. D, 102, 083025,
  \dodoi{10.1103/PhysRevD.102.083025}

\bibitem[{J. Ming {et~al.}(2016)Ming, Krishnan, Papa, Aulbert, \&
  Fehrmann}]{Ming2016}
Ming, J., Krishnan, B., Papa, M.~A., Aulbert, C., \& Fehrmann, H. 2016,
  \bibinfo{title}{Optimal directed searches for continuous gravitational
  waves,} Phys. Rev. D, 93, 064011, \dodoi{10.1103/PhysRevD.93.064011}

\bibitem[{J. {Ming} {et~al.}(2024){Ming}, {Papa}, {Eggenstein}, {Beheshtipour},
  {Machenschalk}, {Prix}, {Allen}, \& {Bensch}}]{ming2024a}
{Ming}, J., {Papa}, M.~A., {Eggenstein}, H.~B., {et~al.} 2024,
  \bibinfo{title}{{Deep Einstein@Home Search for Continuous Gravitational Waves
  from the Central Compact Objects in the Supernova Remnants Vela Jr. and
  G347.3-0.5 Using LIGO Public Data},} \apj, 977, 154,
  \dodoi{10.3847/1538-4357/ad8b9e}

\bibitem[{J. Ming {et~al.}(2022)Ming, Papa, Eggenstein, Machenschalk, Steltner,
  Prix, Allen, \& Behnke}]{Ming2022}
Ming, J., Papa, M.~A., Eggenstein, H.-B., {et~al.} 2022,
  \bibinfo{title}{{Results From an Einstein@Home Search for Continuous
  Gravitational Waves From G347.3 at Low Frequencies in LIGO O2 Data},} The
  Astrophysical Journal, 925, 8, \dodoi{10.3847/1538-4357/ac35cb}

\bibitem[{J. Ming {et~al.}(2018)Ming, Papa, Krishnan, Prix, Beer, Zhu,
  Eggenstein, Bock, \& Machenschalk}]{Ming2018}
Ming, J., Papa, M.~A., Krishnan, B., {et~al.} 2018, \bibinfo{title}{Optimally
  setting up directed searches for continuous gravitational waves in Advanced
  LIGO O1 data,} Phys. Rev. D, 97, 024051, \dodoi{10.1103/PhysRevD.97.024051}

\bibitem[{J. Ming {et~al.}(2019)Ming, Papa, Singh, Eggenstein, Zhu, Dergachev,
  Hu, Prix, Machenschalk, Beer, Behnke, \& Allen}]{Ming2019}
Ming, J., Papa, M.~A., Singh, A., {et~al.} 2019, \bibinfo{title}{Results from
  an Einstein@Home search for continuous gravitational waves from Cassiopeia A,
  Vela Jr., and G347.3,} Phys. Rev. D, 100, 024063,
  \dodoi{10.1103/PhysRevD.100.024063}

\bibitem[{J. Ming {et~al.}(2025)Ming {et~al.}}]{AEIULurl}
Ming, J., {et~al.} 2025,
  \url{www.aei.mpg.de/continuouswaves/o2_casa_velajr_deep-directedsearches}

\bibitem[{L. Mirasola {et~al.}(2024)Mirasola {et~al.}}]{Mirasola:2024kll}
Mirasola, L., {et~al.} 2024, \bibinfo{title}{{New semicoherent targeted search
  for continuous gravitational waves from pulsars in binary systems},} Phys.
  Rev. D, 110, 123043, \dodoi{10.1103/PhysRevD.110.123043}

\bibitem[{L. Mirasola {et~al.}(2025)Mirasola {et~al.}}]{Mirasola:2025car}
Mirasola, L., {et~al.} 2025, \bibinfo{title}{{Search for continuous
  gravitational wave signals from luminous dark photon superradiance clouds
  with LVK O3 observations},} \doarXiv{2501.02052}

\bibitem[{J.~A. Morales \& C.~J. Horowitz(2022)Morales \&
  Horowitz}]{Morales2022}
Morales, J.~A., \& Horowitz, C.~J. 2022, \bibinfo{title}{{Neutron star crust
  can support a large ellipticity},} Monthly Notices of the Royal Astronomical
  Society, 517, 5610, \dodoi{10.1093/mnras/stac3058}

\bibitem[{J.~A. Morales \& C.~J. Horowitz(2024)Morales \&
  Horowitz}]{Morales2024}
Morales, J.~A., \& Horowitz, C.~J. 2024, \bibinfo{title}{Anisotropic neutron
  star crust, solar system mountains, and gravitational waves,} Phys. Rev. D,
  110, 044016, \dodoi{10.1103/PhysRevD.110.044016}

\bibitem[{L. Nieder {et~al.}(2020)Nieder {et~al.}}]{Nieder:2020yqy}
Nieder, L., {et~al.} 2020, \bibinfo{title}{{Discovery of a Gamma-ray Black
  Widow Pulsar by GPU-accelerated Einstein@Home},} Astrophys. J. Lett., 902,
  L46, \dodoi{10.3847/2041-8213/abbc02}

\bibitem[{B.~J. Owen(2010)Owen}]{Owen2010}
Owen, B.~J. 2010, \bibinfo{title}{{How to adapt broad-band gravitational-wave
  searches for $r$-modes},} Phys. Rev. D, 82, 104002,
  \dodoi{10.1103/PhysRevD.82.104002}

\bibitem[{B.~J. Owen {et~al.}(1998)Owen, Lindblom, Cutler, Schutz, Vecchio, \&
  Andersson}]{Owen:1998xg}
Owen, B.~J., Lindblom, L., Cutler, C., {et~al.} 1998,
  \bibinfo{title}{{Gravitational waves from hot young rapidly rotating neutron
  stars},} \dodoi{10.1103/PhysRevD.58.084020}

\bibitem[{B.~J. Owen {et~al.}(2022)Owen, Lindblom, \& Pinheiro}]{Owen2022}
Owen, B.~J., Lindblom, L., \& Pinheiro, L.~S. 2022, \bibinfo{title}{First
  Constraining Upper Limits on Gravitational-wave Emission from NS 1987A in SNR
  1987A,} The Astrophysical Journal Letters, 935, L7,
  \dodoi{10.3847/2041-8213/ac84dc}

\bibitem[{B.~J. Owen {et~al.}(2024)Owen, Lindblom, Pinheiro, \&
  Rajbhandari}]{Owen:2023maw}
Owen, B.~J., Lindblom, L., Pinheiro, L.~S., \& Rajbhandari, B. 2024,
  \bibinfo{title}{{Improved Upper Limits on Gravitational-wave Emission from NS
  1987A in SNR 1987A},} Astrophys. J. Lett., 962, L23,
  \dodoi{10.3847/2041-8213/ad2263}

\bibitem[{M.~A. Papa {et~al.}(2020)Papa, Ming, Gotthelf, Allen, Prix,
  Dergachev, Eggenstein, Singh, \& Zhu}]{Papa_2020midth}
Papa, M.~A., Ming, J., Gotthelf, E.~V., {et~al.} 2020, \bibinfo{title}{Search
  for Continuous Gravitational Waves from the Central Compact Objects in
  Supernova Remnants Cassiopeia A, Vela Jr., and G347.3{\textendash}0.5,} The
  Astrophysical Journal, 897, 22, \dodoi{10.3847/1538-4357/ab92a6}

\bibitem[{G.~G. Pavlov {et~al.}(2001)Pavlov, Sanwal, Kızıltan, \&
  Garmire}]{Pavlov2001}
Pavlov, G.~G., Sanwal, D., Kızıltan, B., \& Garmire, G.~P. 2001,
  \bibinfo{title}{The Compact Central Object in the RX J0852.0–4622 Supernova
  Remnant,} The Astrophysical Journal, 559, L131, \dodoi{10.1086/323975}

\bibitem[{O.~J. Piccinni {et~al.}(2020)Piccinni, Astone, D'Antonio, Frasca,
  Intini, La~Rosa, Leaci, Mastrogiovanni, Miller, \& Palomba}]{Piccinni2020}
Piccinni, O.~J., Astone, P., D'Antonio, S., {et~al.} 2020,
  \bibinfo{title}{Directed search for continuous gravitational-wave signals
  from the Galactic Center in the Advanced LIGO second observing run,} Phys.
  Rev. D, 101, 082004, \dodoi{10.1103/PhysRevD.101.082004}

\bibitem[{H.~J. Pletsch(2008)Pletsch}]{Pletsch2008}
Pletsch, H.~J. 2008, \bibinfo{title}{Parameter-space correlations of the
  optimal statistic for continuous gravitational-wave detection,} Phys. Rev. D,
  78, 102005, \dodoi{10.1103/PhysRevD.78.102005}

\bibitem[{H.~J. Pletsch(2010)Pletsch}]{Pletsch2010}
Pletsch, H.~J. 2010, \bibinfo{title}{Parameter-space metric of semicoherent
  searches for continuous gravitational waves,} Phys. Rev. D, 82, 042002,
  \dodoi{10.1103/PhysRevD.82.042002}

\bibitem[{H.~J. Pletsch \& B. Allen(2009)Pletsch \& Allen}]{Pletsch2009}
Pletsch, H.~J., \& Allen, B. 2009, \bibinfo{title}{Exploiting Large-Scale
  Correlations to Detect Continuous Gravitational Waves,} Phys. Rev. Lett.,
  103, 181102, \dodoi{10.1103/PhysRevLett.103.181102}

\bibitem[{B. Rajbhandari {et~al.}(2021)Rajbhandari, Owen, Caride, \&
  Inta}]{Rajbhandari2021}
Rajbhandari, B., Owen, B.~J., Caride, S., \& Inta, R. 2021,
  \bibinfo{title}{First searches for gravitational waves from $r$-modes of the
  Crab pulsar,} Phys. Rev. D, 104, 122008, \dodoi{10.1103/PhysRevD.104.122008}

\bibitem[{J.~E. {Reed} {et~al.}(1995){Reed}, {Hester}, {Fabian}, \&
  {Winkler}}]{Reed1995}
{Reed}, J.~E., {Hester}, J.~J., {Fabian}, A.~C., \& {Winkler}, P.~F. 1995,
  \bibinfo{title}{{The Three-dimensional Structure of the Cassiopeia A
  Supernova Remnant. I. The Spherical Shell},} \apj, 440, 706,
  \dodoi{10.1086/175308}

\bibitem[{A. Singh \& M.~A. Papa(2023)Singh \& Papa}]{Singh:2022hfd}
Singh, A., \& Papa, M.~A. 2023, \bibinfo{title}{{Opportunistic Search for
  Continuous Gravitational Waves from Compact Objects in Long-period
  Binaries},} Astrophys. J., 943, 99, \dodoi{10.3847/1538-4357/acaf80}

\bibitem[{A. Singh {et~al.}(2017)Singh, Papa, Eggenstein, \&
  Walsh}]{Singh:2017kss}
Singh, A., Papa, M.~A., Eggenstein, H.-B., \& Walsh, S. 2017,
  \bibinfo{title}{{Adaptive clustering procedure for continuous gravitational
  wave searches},} Phys. Rev. D, 96, 082003, \dodoi{10.1103/PhysRevD.96.082003}

\bibitem[{B. Steltner {et~al.}(2022{\natexlab{a}})Steltner, Menne, Papa, \&
  Eggenstein}]{den_cluster}
Steltner, B., Menne, T., Papa, M.~A., \& Eggenstein, H.-B. 2022{\natexlab{a}},
  \bibinfo{title}{Density-clustering of continuous gravitational wave
  candidates from large surveys,} Phys. Rev. D, 106, 104063,
  \dodoi{10.1103/PhysRevD.106.104063}

\bibitem[{B. Steltner {et~al.}(2022{\natexlab{b}})Steltner, Papa, \&
  Eggenstein}]{Steltner2022}
Steltner, B., Papa, M.~A., \& Eggenstein, H.-B. 2022{\natexlab{b}},
  \bibinfo{title}{Identification and removal of non-Gaussian noise transients
  for gravitational-wave searches,} Phys. Rev. D, 105, 022005,
  \dodoi{10.1103/PhysRevD.105.022005}

\bibitem[{B. Steltner {et~al.}(2023)Steltner, Papa, Eggenstein, Prix, Bensch,
  Allen, \& Machenschalk}]{Steltner2023}
Steltner, B., Papa, M.~A., Eggenstein, H.-B., {et~al.} 2023,
  \bibinfo{title}{Deep Einstein@Home All-sky Search for Continuous
  Gravitational Waves in LIGO O3 Public Data,} The Astrophysical Journal, 952,
  55, \dodoi{10.3847/1538-4357/acdad4}

\bibitem[{L. Sun {et~al.}(2018)Sun, Melatos, Suvorova, Moran, \&
  Evans}]{Sun:2017zge}
Sun, L., Melatos, A., Suvorova, S., Moran, W., \& Evans, R.~J. 2018,
  \bibinfo{title}{Hidden Markov model tracking of continuous gravitational
  waves from young supernova remnants,} Phys. Rev. D, 97, 043013,
  \dodoi{10.1103/PhysRevD.97.043013}

\bibitem[{H. {Tananbaum}(1999){Tananbaum}}]{Tanabaum1999}
{Tananbaum}, H. 1999, \bibinfo{title}{{Cassiopeia A},} IAU Cirulars, 7246, 1

\bibitem[{G. Ushomirsky {et~al.}(2000)Ushomirsky, Cutler, \&
  Bildsten}]{Ushomirsky2000}
Ushomirsky, G., Cutler, C., \& Bildsten, L. 2000, \bibinfo{title}{{Deformations
  of Accreting Neutron Star Crusts and Gravitational Wave Emission},} Mon.
  Notices Royal Astron. Soc., 319, 902,
  \dodoi{10.1046/j.1365-8711.2000.03938.x}

\bibitem[{A.~F. Vargas \& A. Melatos(2024)Vargas \& Melatos}]{Vargas:2024itq}
Vargas, A.~F., \& Melatos, A. 2024, \bibinfo{title}{{Stochastic and secular
  anomalies in pulsar braking indices},} Mon. Not. Roy. Astron. Soc., 534,
  3410, \dodoi{10.1093/mnras/stae2326}

\bibitem[{J. Wang \& K. Riles(2024)Wang \& Riles}]{PhysRevD.110.042006}
Wang, J., \& Riles, K. 2024, \bibinfo{title}{Deep search of the full O3 LIGO
  data for continuous gravitational waves from the Cassiopeia A central compact
  object,} Phys. Rev. D, 110, 042006, \dodoi{10.1103/PhysRevD.110.042006}

\bibitem[{K. {Wette} {et~al.}(2008){Wette}, {Owen}, {Allen}, {Ashley},
  {Betzwieser}, {Christensen}, {Creighton}, {Dergachev}, {Gholami}, {Goetz},
  {Gustafson}, {Hammer}, {Jones}, {Krishnan}, {Landry}, {Machenschalk},
  {McClelland}, {Mendell}, {Messenger}, {Papa}, {Patel}, {Pitkin}, {Pletsch},
  {Prix}, {Riles}, {Sancho de la Jordana}, {Scott}, {Sintes}, {Trias},
  {Whelan}, \& {Woan}}]{wette2008}
{Wette}, K., {Owen}, B.~J., {Allen}, B., {et~al.} 2008,
  \bibinfo{title}{{Searching for gravitational waves from Cassiopeia A with
  LIGO},} Classical and Quantum Gravity, 25, 235011,
  \dodoi{10.1088/0264-9381/25/23/235011}

\bibitem[{M. {Yu} {et~al.}(2013){Yu}, {Manchester}, {Hobbs}, {Johnston},
  {Kaspi}, {Keith}, {Lyne}, {Qiao}, {Ravi}, {Sarkissian}, {Shannon}, \&
  {Xu}}]{glitch_2013}
{Yu}, M., {Manchester}, R.~N., {Hobbs}, G., {et~al.} 2013,
  \bibinfo{title}{{Detection of 107 glitches in 36 southern pulsars},} MNRAS,
  429, 688, \dodoi{10.1093/mnras/sts366}

\bibitem[{Y. Zhang {et~al.}(2021)Zhang, Papa, Krishnan, \& Watts}]{Zhang2021}
Zhang, Y., Papa, M.~A., Krishnan, B., \& Watts, A.~L. 2021,
  \bibinfo{title}{Search for Continuous Gravitational Waves from Scorpius X-1
  in LIGO O2 Data,} The Astrophysical Journal Letters, 906, L14,
  \dodoi{10.3847/2041-8213/abd256}

\bibitem[{S. {Zhou} {et~al.}(2022){Zhou}, {G{\"u}gercino{\u{g}}lu}, {Yuan},
  {Ge}, \& {Yu}}]{glitch_rev_2022}
{Zhou}, S., {G{\"u}gercino{\u{g}}lu}, E., {Yuan}, J., {Ge}, M., \& {Yu}, C.
  2022, \bibinfo{title}{{Pulsar Glitches: A Review},} Universe, 8, 641,
  \dodoi{10.3390/universe8120641}

\bibitem[{S.~J. Zhu {et~al.}(2020)Zhu, Baryakhtar, Papa, Tsuna, Kawanaka, \&
  Eggenstein}]{Zhu2020}
Zhu, S.~J., Baryakhtar, M., Papa, M.~A., {et~al.} 2020,
  \bibinfo{title}{Characterizing the continuous gravitational-wave signal from
  boson clouds around Galactic isolated black holes,} Phys. Rev. D, 102,
  063020, \dodoi{10.1103/PhysRevD.102.063020}

\bibitem[{M. Zimmermann \& E. Szedenits(1979)Zimmermann \&
  Szedenits}]{zimmermann:1979}
Zimmermann, M., \& Szedenits, E. 1979, \bibinfo{title}{Gravitational waves from
  rotating and precessing rigid bodies: Simple models and applications to
  pulsars,} Phys. Rev. D, 20, 351, \dodoi{10.1103/PhysRevD.20.351}

\end{thebibliography}
\bibliographystyle{aasjournal}

\end{document}